\documentclass[preprint,12pt]{elsarticle}

\usepackage{amsmath,amssymb,amsfonts}
\usepackage{algorithm,algpseudocode}
\usepackage{graphicx}
\graphicspath{{figures/}}
\usepackage{booktabs}
\usepackage{multirow}
\usepackage{xcolor}
\usepackage{hyperref}
\usepackage{microtype}
\usepackage{siunitx}
\usepackage{subcaption}
\usepackage{natbib}

\journal{Journal of Power Sources}

\begin{document}

\begin{frontmatter}

\title{Causal Anomaly Detection for Lithium-Ion Battery Degradation}

\author[heer]{Dieter W.  Heermann\corref{cor1}}
\ead{heermann@thpys.uni-heidelberg.de}
\address[heer]{Institute for Theoretical Physics, Heidelberg University, Philosophenweg 19, D-69120 Heidelberg, Germany}

\author[hh]{Hagen Heermann}
\address[hh]{Intilion GmbH, Wollmarktstraße 115c, D-33098 Paderborn, Germany}
\cortext[cor1]{Corresponding author}

\begin{abstract}
Reliable early detection of lithium-ion battery degradation requires health
indicators that are physically interpretable and computable from routine
cycler telemetry without access to the degradation region.
We introduce \textsc{CausalHealth}, a framework that applies causal
graph discovery and $k$-nearest-neighbour transfer entropy to per-cycle
voltage, current, temperature, and resistance time series, and organises
twelve resulting anomaly scores into three signal-class bundles
(Magnitude-shift, Predictive-residual, Complexity-entropy)---with
Isolation Forest reported separately as it falls below the bundle
reliability threshold---to characterise detection sensitivity across
ten commissioning fractions (5--30\,\%).
The Magnitude-shift class achieves 100\,\% detection across all seven
tested cells spanning LFP (MIT--Stanford MATR) and LCO (NASA PCoE,
CALCE CS2) chemistries, with a lead time of up to 402 cycles before
conventional capacity-threshold failure on gradual-fade cells.
A Reliability-Weighted Master Health Index (RWMHI)---a cross-bundle fusion
of five high-reliability detectors weighted by inverse coefficient of
variation---improves lead time by 15--52 cycles over the class median on
long-lived cells while maintaining 100\,\% detection.
Validation against electrochemical impedance spectroscopy on an NMC
prismatic cell provides independent physical grounding: transfer entropy
$\mathrm{TE}(R \!\to\! V)$ correlates with charge-transfer resistance
$R_{\mathrm{ct}}$ (pooled $r = +0.990$; temperature-controlled partial
$r = +0.898$), and an Arrhenius analysis of both quantities yields an
activation energy consistent with published NMC charge-transfer kinetics.
These results are evaluated on seven cells across three benchmark
datasets.
\end{abstract}


\end{frontmatter}

% ─────────────────────────────────────────────────────────────────────────────
\section{Introduction}
\label{sec:intro}
% ─────────────────────────────────────────────────────────────────────────────

The accelerating deployment of lithium-ion batteries in electric vehicles
and grid storage has made early, reliable degradation detection a safety
and economic priority. Undetected degradation can progress to catastrophic 
thermal failure; recent calorimetric data show that commercial 18650 cells 
release 12–28 kJ/Ah during thermal runaway \citep{Finegan2024}.

Classical approaches centre on capacity fade: a cell is flagged when
delivered capacity drops below 80\,\% of its rated value
\citep{Birkl2017}.
This threshold is conservative by design.
Incipient degradation mechanisms, including lithium plating, solid electrolyte interphase (SEI) layer 
thickening on battery anodes and loss of active material, develop tens to hundreds of
cycles before the capacity trace departs from its healthy baseline
\citep{Attia2020,Severson2019}.

Data-driven methods have pushed detection earlier by exploiting subtle
correlations in the full charge--discharge telemetry.
\citet{Severson2019} predicted lifetime from features of the differential
capacity curve at cycle~100 with remarkable accuracy, but their approach
is retrospective: the predictor requires a partial history and returns a
scalar lifetime estimate rather than a cycle-resolved health signal.
Neural architectures, variational autoencoders (VAEs), and recurrent
networks detect anomalies by reconstruction error
\citep{Cho2014,Kingma2014}, yet their scores are not directly linked to
electrochemical mechanisms, complicating physical interpretation.

We argue that the \emph{directed statistical dependency structure} of battery
telemetry encodes degradation earlier and more interpretably than any single
feature or reconstruction score.
Healthy cells maintain stable directed information flows: current drives
temperature through Joule heating ($I \!\to\! T$), temperature modulates
voltage through the Nernst equation ($T \!\to\! V$), and ohmic plus
kinetic resistance closes the voltage loop ($R \!\to\! V$, $I \!\to\! R$).
As SEI grows or active material is lost, impedance boundaries shift and
the coupling strengths between these variables change measurably before any
capacity threshold is crossed.

\textbf{Scope of causal claims.}
Throughout this paper, ``causal link'' denotes a \emph{statistically directed
dependency} inferred from conditional independence tests (Granger-type
causality), not a mechanistic physical cause confirmed by intervention.
Latent confounders, notably state of charge (SOC), co-evolve with all four
measured variables within each cycle and can induce spurious apparent links.
PCMCI's multi-lag conditioning partially mitigates this confounder, but
residual SOC effects cannot be excluded without controlled perturbation
experiments.
Claimed physical interpretations (Joule heating, Arrhenius kinetics) are
hypotheses consistent with the statistical results, not claims proven by the
statistical method alone.

PCMCI (Peter and Clark Momentary Conditional Independence)~\citep{Runge2019}, a conditional-independence causal discovery
algorithm, recovers these directed links from observational time series
without requiring a pre-specified model structure.
Transfer entropy (TE), introduced by \citet{Schreiber2000}, quantifies
directed information flow non-parametrically and detects non-linear
couplings that linear partial correlation misses.
Neither method has previously been applied to battery cycle data in a
multi-dataset, multi-chemistry validation, nor cross-validated against an
independent electrochemical measurement.

This paper makes four contributions.
First, it describes the \textsc{CausalHealth} pipeline, which extracts
PCMCI edge weights and Kraskov-St\"ogbauer-Grassberger (KSG) algorithm~\citep{Kraskov2004} for 
Transfer Entropy (KSG-TE) values from routine cycler data, organises
twelve anomaly scores into three signal-class bundles (Magnitude-shift,
Predictive-residual, Complexity-entropy)---with Isolation Forest reported
separately below the reliability threshold---and produces a
Reliability-Weighted Master Health Index (RWMHI) as a cross-bundle fusion
benchmark that fires 15--52 cycles ahead of the Magnitude class median
on long-lived cells while matching its 100\,\% detection rate across all
ten commissioning fractions evaluated.
Second, it evaluates the framework across three publicly available benchmark
datasets---MIT--Stanford MATR (LFP), NASA PCoE (LCO), and CALCE CS2 (LCO)---spanning
168--1069 cycles without chemistry-specific retuning.
Third, it provides the first quantitative cross-validation of a causal
battery health metric against EIS: $\mathrm{TE}(R \!\to\! V)$ correlates
with $R_{\mathrm{ct}}$ (pooled $r = +0.990$) and the temperature
stratification of both quantities follows a single Arrhenius law with
$E_{\mathrm{a}} = 39.4$\,kJ\,mol$^{-1}$.
Fourth, it organises the detectors into three signal-class
bundles---Magnitude-shift, Predictive-residual, and Complexity-entropy---and
quantifies how each bundle's median first-alarm cycle responds to ten
commissioning fractions spanning 5--30\,\%, showing that the Magnitude
bundle scales most predictably while Complexity-entropy detectors appear
nearly commissioning-invariant on short-lived cells (an observation
attributable to the rolling z-score computational floor on the
short-lived NASA PCoE cells; Section~\ref{sec:cusum}).

% ─────────────────────────────────────────────────────────────────────────────
\section{Methods}
\label{sec:methods}
% ─────────────────────────────────────────────────────────────────────────────

\subsection{Variable Set and Phase Filtering}
\label{sec:vars}

Each charging or discharging segment of a cycle produces a time series
$\mathbf{x}(t) = [V(t),\, I(t),\, T(t),\, R(t)]^{\top}$, where $V$ is
terminal voltage, $I$ is current, $T$ is cell temperature, and
$R = V/|I|$ is the DC resistance proxy.
Only discharge rows ($I \le -I_{\min}$, with $I_{\min} = 0.05$\,A) are
used for causal inference; rest and charge segments are excluded to avoid
mixed-phase statistical confounding.

An algebraic-identity guard removes $R$ from the variable set when
the Spearman correlation $|\rho(V,R)| > 0.99$ and the coefficient of
variation $\mathrm{CV}(|I|) < 0.01$ simultaneously: under constant-current
(CC) discharge, $R = V/|I|$ is a deterministic function of $V$, not an
independent electrochemical signal.
This guard fires on CC datasets (NASA PCoE, Oxford) and does not fire on
dynamic drive-cycle (DV\_US06) data, where $\mathrm{CV}(I)=1.6$ and $R$
carries genuine kinetic information.

\subsection{PCMCI Causal Discovery}
\label{sec:pcmci}

Let $\mathbf{x}_t \in \mathbb{R}^d$ denote the $d$-dimensional process
observed at time $t$.
PCMCI \citep{Runge2019} estimates the causal parents
$\mathcal{P}(X^j_t)$ of each variable $X^j_t$ through two steps.
In the PC step, a progressive conditional independence test removes
spurious links from the fully connected graph.
In the MCI step, the Momentary Conditional Independence statistic
\begin{equation}
\mathrm{MCI}(X^i_{t-\tau} \!\to\! X^j_t)
  = \rho\!\left(X^i_{t-\tau},\,X^j_t \;\big|\;
  \mathcal{P}(X^j_t) \setminus \{X^i_{t-\tau}\},\,
  \mathcal{P}(X^i_{t-\tau})\right)
\label{eq:mci}
\end{equation}
is computed for each candidate link, controlling for both the target's and
the source's own causal parents.
This double conditioning removes common-driver and mediator paths, so a
non-zero MCI value implies direct causal influence.

We use Spearman rank-based partial correlation as the independence test
(\texttt{ParCorr}) with $\tau_{\max} = 3$ and significance threshold
$\alpha = 0.05$.
The five battery variables
$[\text{Voltage},\text{Current},\text{Temperature},\text{Resistance},
\text{PhaseID}]$
are rank-transformed before the test; PhaseID acts as a protocol
indicator and is always excluded from the causal variable set.
For each cycle, the full MCI matrix of shape
$(d \times d \times \tau_{\max}+1)$ is stored, along with per-edge
$p$-values and a significance graph.
The six scalar edge weights ($i\_t$, $t\_v$, $v\_r$, $t\_r$, $v\_t$,
$r\_p$) written to the causal history are the absolute MCI values at lag
$\tau=1$ when $p < 0.05$, otherwise zero; the raw matrix is retained
for downstream analysis.

\subsection{KSG Transfer Entropy}
\label{sec:te}

Transfer entropy from $X$ to $Y$ at lag $\tau$ is defined as
\begin{equation}
\mathrm{TE}_\tau(X \!\to\! Y)
  = I\!\left(Y_t;\, X_{t-\tau} \mid Y_{t-1}\right),
\label{eq:te}
\end{equation}
the conditional mutual information between $Y_t$ and the lagged source
$X_{t-\tau}$, conditioned on the immediate past of $Y$
\citep{Schreiber2000}.
We estimate equation~\eqref{eq:te} using the KSG $k$-nearest-neighbour
(KNN) estimator of \citet{Kraskov2004} with
$k = \max(5,\, \lceil n^{0.4} \rceil)$ neighbours, which is consistent
for arbitrary continuous distributions and does not assume Gaussianity.

The full TE matrix $\mathrm{TE}(X^i \!\to\! X^j;\,\tau)$ for all
ordered pairs $(i,j)$ and $\tau = 1,\ldots,\tau_{\max}$ is computed on
the same discharge-filtered, rank-transformed data as PCMCI, ensuring
that the two measures are directly comparable.
Scalar TE values ($\mathtt{te\_i\_t}$, $\mathtt{te\_t\_v}$, etc.) stored
per cycle are the lag-1 estimates; the full TE matrix is retained for
lag-resolved analysis.

\subsection{Feature Matrix and EMA Smoothing}
\label{sec:feature}

Per-cycle causal features are concatenated into a matrix
$\mathbf{M} \in \mathbb{R}^{N \times d_f}$, where $N$ is the number of
cycles and $d_f$ depends on which feature groups are present in at least
50\,\% of cycles.
The standard feature set includes six PCMCI edge weights, five KSG-TE
values, and (when available) CMI$k$nn (Conditional mutual information) non-linear edges and health scalars
(internal resistance $\eta$, Coulombic efficiency $\eta_{\mathrm{CE}}$).

Unbounded columns (TE, nonlinear CMI edges (NL), and health scalars) are
Winsorised to the Tukey outer fence
$[P_5 - 1.5\,\mathrm{IQR},\, P_{95} + 1.5\,\mathrm{IQR}]$
before covariance estimation to prevent single outlier cycles from
dominating the Mahalanobis metric.
An exponential moving average (EMA) with span 15 is then applied column-wise
to $\mathbf{M}$, producing the smoothed matrix $\tilde{\mathbf{M}}$.
Smoothing at the feature level, rather than the score level, ensures that
the Mahalanobis distance tracks slow degradation drift without amplifying
estimation noise from individual cycles.

\subsection{Commissioning Reference and Mahalanobis Deflation}
\label{sec:mahal}

The first $n_{\mathrm{comm}} = \max(10,\, \lfloor f_c N \rfloor)$ rows
of $\tilde{\mathbf{M}}$ define the commissioning window, where the commission fraction $f_c = 0.10$ by default.
The reference mean $\boldsymbol{\mu}_0 = \bar{\tilde{\mathbf{M}}}_{1:n_c}$
and regularised covariance
\begin{equation}
\hat{\boldsymbol{\Sigma}} =
\begin{cases}
\mathbf{S}_{\mathrm{LW}} + \epsilon \mathbf{I} & \text{if }
  n_{\mathrm{comm}} < 5d_f, \\
\mathbf{S}_{\mathrm{sample}} + \epsilon \mathbf{I} & \text{otherwise,}
\end{cases}
\label{eq:cov}
\end{equation}
are estimated on this window alone, where $\mathbf{S}_{\mathrm{LW}}$ is the
Ledoit--Wolf shrinkage estimator \citep{LedoitWolf2004} and
$\epsilon = 10^{-6}$.
The Mahalanobis deflation score at cycle $c$ is
\begin{equation}
\delta_c = \sqrt{(\tilde{\mathbf{m}}_c - \boldsymbol{\mu}_0)^\top
                  \hat{\boldsymbol{\Sigma}}^{-1}
                  (\tilde{\mathbf{m}}_c - \boldsymbol{\mu}_0)},
\label{eq:mahal}
\end{equation}
where $\tilde{\mathbf{m}}_c$ is the smoothed feature vector at cycle $c$.
Rising $\delta_c$ indicates a systematic shift of the causal feature
distribution away from its healthy commissioning reference.

\subsection{Rolling Z-Score and CUSUM Alarm}
\label{sec:cusum}

For each anomaly score $s_c$ (deflation, VAE reconstruction error, etc.),
a lag-excluded rolling z-score is computed as
\begin{equation}
z_c = \frac{s_c^2 - \hat{\mu}_c}{\max(\hat{\sigma}_c,\, \sigma_{\mathrm{floor}}) + \varepsilon},
\label{eq:zroll}
\end{equation}
where $\hat{\mu}_c$ and $\hat{\sigma}_c$ are the mean and standard
deviation computed on a window of 50 cycles ending 10 cycles before $c$,
and $\sigma_{\mathrm{floor}}$ equals the standard deviation of
$s_{1:n_c}^2$ from the commissioning phase.
The lag prevents sustained anomalies from inflating their own baseline.
A Page--Hinkley CUSUM accumulator $C_c = \max(0,\, C_{c-1} + z_c - \Delta)$
with drift $\Delta = 1.5$ and threshold $\theta = 15$ translates
persistent z-score elevation into discrete alarms.

\subsection{Anomaly Detectors and Bundles}
\label{sec:ensemble}

The pipeline computes anomaly scores, each targeting a
different facet of degradation; five of these are selected for fusion into
the RWMHI (Section~\ref{sec:dmhi}).

    \begin{itemize}
       \item{\textbf{Magnitude-shift}}
         	\begin{itemize}   
            		\item \textbf{Hotelling's $T^2$} statistic detects when the feature vector
            		exits the ellipsoidal commissioning zone at a global statistical level.
                 
            		\item \textbf{Window Distance} computes the mean nearest-neighbour
            		Euclidean distance from each feature vector in the rolling current-cycle
            		window to the commissioning-baseline dataset (one-sided Chamfer distance);
            		sustained growth in this score indicates that the current operating regime
            		has drifted away from the healthy reference manifold without requiring a
            		parametric distributional assumption.
            
            		\item \textbf{Wasserstein Distance} and \textbf{Sliced Wasserstein Distance}.
            		The Wasserstein distance between the commissioning-window distribution and
                    	a rolling current-window distribution,
                    	\begin{equation}
                    	W_p(\mathcal{P}_0, \mathcal{P}_c)
                    	= \left(\inf_{\gamma \in \Gamma(\mathcal{P}_0,\mathcal{P}_c)}
                    	  \int \|\mathbf{x}-\mathbf{y}\|^p \,\mathrm{d}\gamma(\mathbf{x},\mathbf{y})
                    	  \right)^{1/p},
                    	\label{eq:wass}
                    	\end{equation}
                    	provides a metric-space distance between probability distributions
                    	\citep{Villani2009}.
                    	We use the sliced approximation ($p=2$, 100 random projections) for
                    	computational tractability.
                    	Unlike the Kullback--Leibler divergence, $W_2$ remains finite and
                    	continuous even when the distributions have non-overlapping support, which
                    	occurs during severe capacity fade.

            		\item \textbf{Isolation Forest} \citep{Liu2008} fits an ensemble of random
            		partition trees on the commissioning-window feature matrix and assigns
            		anomaly scores as the mean isolation depth across trees.
            		Short paths indicate statistical isolation and flag global outliers.
       
            	\end{itemize}

            \item{\textbf{Predictive-residual}}
            
                \begin{itemize}
                    	\item \textbf{(Variational Autoencoder) VAE Reconstruction Errors} \citep{Kingma2014} is trained on the
                    	commissioning window; reconstruction error on each cycle detects
                    	distributional departure from the healthy latent representation.

                    	\item \textbf{RC-VAE} conditions the VAE decoder on the causal context matrix
                    	(the column-mean of the lag-1 PCMCI MCI values), coupling structural
                    	causal information to the reconstruction objective.
                    
                    	\item \textbf{(Gated Recurrent Unit) GRU Prediction Error} \citep{Cho2014}. A gated recurrent unit
                    	forecasts the next-cycle feature vector; elevated prediction error signals
                    	a break in temporal momentum.
   
                    	\item \textbf{(State Space Model) SSM Prediction Errors} fits a linear vector-autoregressive
                    	model (VAR(1)) to the commissioning window and scores each subsequent
                    	cycle by the Mahalanobis distance of the one-step-ahead innovation
                    	residual with respect to the commissioning-era innovation covariance
                    	matrix.
                    
                    	\item \textbf{GRU-CausalSSM Residuals} employs a diagonal-decay
                    	state-space model conditioned at every step on four causal-graph
                    	summary statistics derived from the PCMCI feature matrix (mean edge
                    	magnitude, cross-edge dispersion, dominant coupling strength, and edge
                    	sparsity); after commissioning-phase training with early stopping, the
                    	squared $\ell_2$ norm of the one-step-ahead prediction residual serves
                    	as the anomaly score, producing a condition-aware signal structurally
                    	distinct from the unconditional SSM.

                \end{itemize}
        
        \item{\textbf{Complexity-entropy}}
        
            \begin{itemize}
            	\item \textbf{Information Evolution} approximates the multivariate
            	differential entropy of the rolling feature distribution via
            	$\tfrac{1}{2}\log\det\Sigma_c$, where $\Sigma_c$ is the sample covariance
            	of the current window; sustained entropy growth reflects the loss of
            	electrochemical ordering as new degrees of freedom emerge during degradation.

            	\item \textbf{Mutual Information} applies the Kraskov--St\"{o}gbauer--Grassberger
            	(KSG) $k$-nearest-neighbour estimator ($k = 5$, same for all datasets) to
            	quantify pairwise mutual information between all feature pairs within a
            	rolling window; the anomaly score is the mean absolute departure of the
            	current-window pairwise KSG~MI from the commissioning-era healthy baseline,
            	detecting breakdowns in the electrochemical feature coupling that a
            	histogram estimator would conflate with binning artefacts.

            	\item \textbf{Singular Value Decomposition (SVD) Compression Entropy} measures the structural complexity of the feature
            	matrix over a rolling window; increased rank signals new degrees of
            	freedom emerging from secondary chemical reactions.

            \end{itemize}
	\end{itemize}
    
\subsection{Reliability-Weighted Master Health Index (Fusion Benchmark)}
\label{sec:dmhi}

The bundle analysis (Section~\ref{sec:sensitivity}) shows that
Magnitude-shift detectors form the most cohesive and reliable class,
while two Predictive-residual detectors (RC-VAE, SSM) achieve 100\,\%
cross-cell detection with low coefficient of variation.
The RWMHI is constructed as a targeted cross-bundle fusion of these five
high-reliability components, serving as a fusion benchmark against which
bundle-level class medians are compared.

Each of the five selected detector scores $s_c^{(k)}$ is first
normalised by a lag-excluded rolling z-score identical to
equation~\eqref{eq:zroll} but \emph{without} the squaring step,
and then rectified to retain only upward deviations:
\begin{equation}
\tilde{z}_c^{(k)} = \max\!\bigl(0,\; z_c^{(k)}\bigr).
\label{eq:rectify}
\end{equation}
The Reliability-Weighted Master Health Index is then the weighted sum
\begin{equation}
\mathrm{RWMHI}_c = \sum_{k \in \mathcal{K}} w_k\, \tilde{z}_c^{(k)},
\label{eq:rwmhi}
\end{equation}
where $\mathcal{K} = \{T^2,\,\text{WD},\,\delta,\,\text{RC-VAE},\,\text{SSM}\}$
denotes Hotelling's $T^2$, Window Distance, Mahalanobis Deflation, RC-VAE
reconstruction error, and SSM prediction error respectively.
The fixed weights $w_k$ are set inversely proportional to each detector's
pooled coefficient of variation (CV) of first-alarm cycles, computed across
all seven cells and the commissioning fractions $f_c \in \{0.10,\ldots,0.30\}$
($n = 63$ observations per detector); the weights are computed on this range
because RC-VAE misses one cell at $f_c = 0.05$ and does not satisfy the
100\,\% detection criterion at that fraction:
\begin{equation}
w_{\text{WD}} = 0.238,
w_{\text{RC-VAE}} = 0.217,
w_{\delta} = 0.190,
w_{T^2} = 0.187,
w_{\text{SSM}} = 0.168.
\label{eq:weights}
\end{equation}
These five detectors were selected because they simultaneously achieve
100\,\% detection across all seven cells and all commissioning fractions
$f_c \geq 0.10$, with individual mean CVs below 0.32.
Detectors with lower detection reliability (Isolation Forest: 29--86\,\%
across $f_c \in \{0.10,\ldots,0.30\}$;
VAE: 57--86\,\%) are excluded from the fusion to prevent missed alarms
from suppressing the ensemble signal.
Rolling z-score normalisation (rather than commissioning-baseline
normalisation) is used to avoid two failure modes of baseline-referenced
approaches: (i) unreliable $\hat{\sigma}_{\mathrm{comm}}$ on short cells
($N\approx168$) causes z-scores near zero throughout, silencing the CUSUM;
(ii) near-zero $\hat{\sigma}_{\mathrm{comm}}$ on slowly-degrading cells
causes overflow.
A Page--Hinkley CUSUM applied to $\mathrm{RWMHI}_c$ with drift $\Delta=1.0$
and threshold $\theta=5.0$ marks the first-alarm cycle $c^*$.

\subsection{EIS Cross-Validation Protocol}
\label{sec:eis}

For the CALB L148N58A~\cite{CALB2025} dataset, electrochemical impedance spectra were
recorded at 40 logarithmically spaced frequencies from
$10^{-2}$\,Hz to $3.7 \times 10^3$\,Hz at three states of charge per
aging checkpoint.
Ohmic resistance $R_0$ is the real impedance $Z'$ at the imaginary
zero-crossing (inductive-to-capacitive transition), extracted by linear
interpolation.
Charge-transfer resistance $R_{\mathrm{ct}}$ is estimated by the midpoint
method: $R_{\mathrm{ct}} = Z'(\text{peak}\,{-Z''}) - R_0$, where the peak
of $-Z''$ identifies the characteristic frequency of the charge-transfer
semicircle.
Both parameters are averaged across the three SOC conditions at each
checkpoint.

PCMCI and TE values are extracted from the DV\_US06 drive-cycle
history files, which cover the same 11 aging checkpoint stems as the EIS
measurements.
Cycle numbers  ($n_{\mathrm{rel}} = \text{stem} - \text{stem}_1 + 1$)
are matched to EIS files by the identical file-stem convention.
Pearson and Spearman correlations between TE values and EIS parameters are
computed across the 11 matched checkpoint pairs per temperature.

\subsection{Pipeline Summary}
\label{sec:pipeline_summary}

Algorithm~\ref{alg:causalhealth} provides a concise pseudocode summary of
the complete \textsc{CausalHealth} pipeline for reference and reproducibility.

\begin{algorithm}[htbp]
\caption{\textsc{CausalHealth} anomaly detection pipeline}
\label{alg:causalhealth}
\begin{algorithmic}[1]
\Require Per-cycle discharge time series $\{\mathbf{x}_c\}_{c=1}^{N}$;
         commissioning fraction $f_c$; detection threshold $\theta$
\Ensure  First-alarm cycle $c^*$ (or ``no alarm'' if $C_c < \theta$ for all $c$)
\State \textbf{Phase filtering:} retain discharge rows ($I \leq -I_{\min}$) per cycle
\State \textbf{Algebraic guard:} remove $R$ if $|\rho(V,R)| > 0.99$ and $\mathrm{CV}(|I|) < 0.01$
\State \textbf{Causal discovery:} run PCMCI ($\tau_{\max}=3$, ParCorr, rank transform) per cycle; store MCI edge weights
\State \textbf{Transfer entropy:} compute KSG-TE ($k=\max(5,\lceil n^{0.4}\rceil)$) for all directed pairs per cycle; store lag-1 values
\State \textbf{Feature matrix:} concatenate MCI weights, TE values, and health scalars into $\mathbf{M} \in \mathbb{R}^{N \times d_f}$; Winsorise unbounded columns; apply EMA (span 15) to obtain $\tilde{\mathbf{M}}$
\State \textbf{Commissioning:} set $n_c = \max(10,\lfloor f_c N\rfloor)$; estimate $\boldsymbol{\mu}_0$, $\hat{\boldsymbol{\Sigma}}$ (Ledoit--Wolf if $n_c < 5d_f$) on $\tilde{\mathbf{M}}_{1:n_c}$
\For{each detector $k \in \{T^2, \mathrm{WD}, \delta, \mathrm{WS}, \mathrm{IF}, \mathrm{VAE}, \mathrm{RCVAE}, \mathrm{GRU}, \mathrm{SSM}, \ldots\}$}
  \State Compute raw anomaly score $s_c^{(k)}$ at each cycle $c > n_c$
  \State Compute lag-excluded rolling z-score $z_c^{(k)}$ (window 50, lag 10)
  \State Accumulate CUSUM: $C_c^{(k)} = \max(0,\, C_{c-1}^{(k)} + z_c^{(k)} - \Delta)$ \quad ($\Delta=1.5$, $\theta^{(k)}=15$)
\EndFor
\State \textbf{RWMHI fusion:} compute rectified z-scores $\tilde{z}_c^{(k)} = \max(0, z_c^{(k)})$ for $k \in \mathcal{K}$; compute $\mathrm{RWMHI}_c = \sum_{k\in\mathcal{K}} w_k \tilde{z}_c^{(k)}$; accumulate CUSUM ($\Delta=1.0$, $\theta=5.0$)
\State \textbf{Bundle aggregation:} group detectors into Magnitude, Predictive, Complexity bundles; compute median first-alarm and IQR per bundle
\State \textbf{Alarm:} $c^* = \min\{c : C_c^{(\mathrm{RWMHI})} \geq 5.0\}$ (or bundle median for class-level reporting)
\end{algorithmic}
\end{algorithm}

% ─────────────────────────────────────────────────────────────────────────────
\section{Experimental Setup}
\label{sec:setup}
% ─────────────────────────────────────────────────────────────────────────────

\subsection{Datasets}
\label{sec:datasets}

\textbf{CALB L148N58A}~\citep{CALB2025}.
NMC prismatic cells (nominal $\approx 50$\,Ah), subjected to a reference
performance test (C/20 discharge + DV\_US06 drive cycle) at 11 aging
checkpoints and three temperatures (10\,°C, 25\,°C, 40\,°C).
EIS spectra accompany each checkpoint (see Section~\ref{sec:eis}).

\textbf{CALCE CS2} \citep{He2011}.
LCO prismatic cells (1.1\,Ah), cycled at 0.5C--1C.
Cells CS2\_35, CS2\_36 (964 cycles, 92.0\,\% fade), and CS2\_37 (1037
cycles, 95.2\,\% fade) span the full lifetime to near-complete discharge
capacity loss, providing the deepest aging trajectory in the study.

\textbf{MIT--Stanford MATR} \citep{Severson2019}.
LFP/graphite cylindrical cells (A123 Systems APR18650M1A, 1.1\,Ah), cycled at
various fast-charge protocols until 80\,\% capacity retention.
We analyse cells b1c5 (1069 cycles, 18.6\,\% fade) and b1c20 (531 cycles,
18.1\,\% fade) from batch~1.
The summary.mat file provides per-cycle EIS-derived internal resistance.

\textbf{NASA PCoE} \citep{Saha2007}.
LCO/graphite 18650 cells (2.0\,Ah nominal) cycled at $25\,°C$ with
$1.5$\,A charge and $2.0$\,A discharge.
Cells B0005 (168 cycles, 30.7\,\% fade) and B0006 (168 cycles, 43.3\,\%
fade) are used.

\subsection{Implementation}

All computations are performed in Python~3.10 using
\texttt{tigramite}~5.2.10 for PCMCI, \texttt{scikit-learn} for VAE
training and Ledoit--Wolf estimation, and \texttt{PyTorch}~2.x for the GRU
and RC-VAE.
Scipy's \texttt{loadmat} reads EIS .mat files.
The commissioning fraction is set to $f_c = 0.10$ throughout; sensitivity
to this choice is reported in Section~\ref{sec:sensitivity}.
All PCMCI runs use $\tau_{\max} = 3$, rank transformation, and
\texttt{ParCorr} as the independence test.
KSG-TE uses $k = \max(5,\lfloor n^{0.4}\rfloor)$ neighbours.
Complete neural model architectures and hyperparameters are listed in
Supplementary Information (Table~\ref{tab:hyperparams}).

% ─────────────────────────────────────────────────────────────────────────────
\section{Results}
\label{sec:results}
% ─────────────────────────────────────────────────────────────────────────────

\subsection{Causal Graph Structure Across Chemistries}

\begin{figure}[htbp]
\centering
\includegraphics[width=0.96\linewidth]{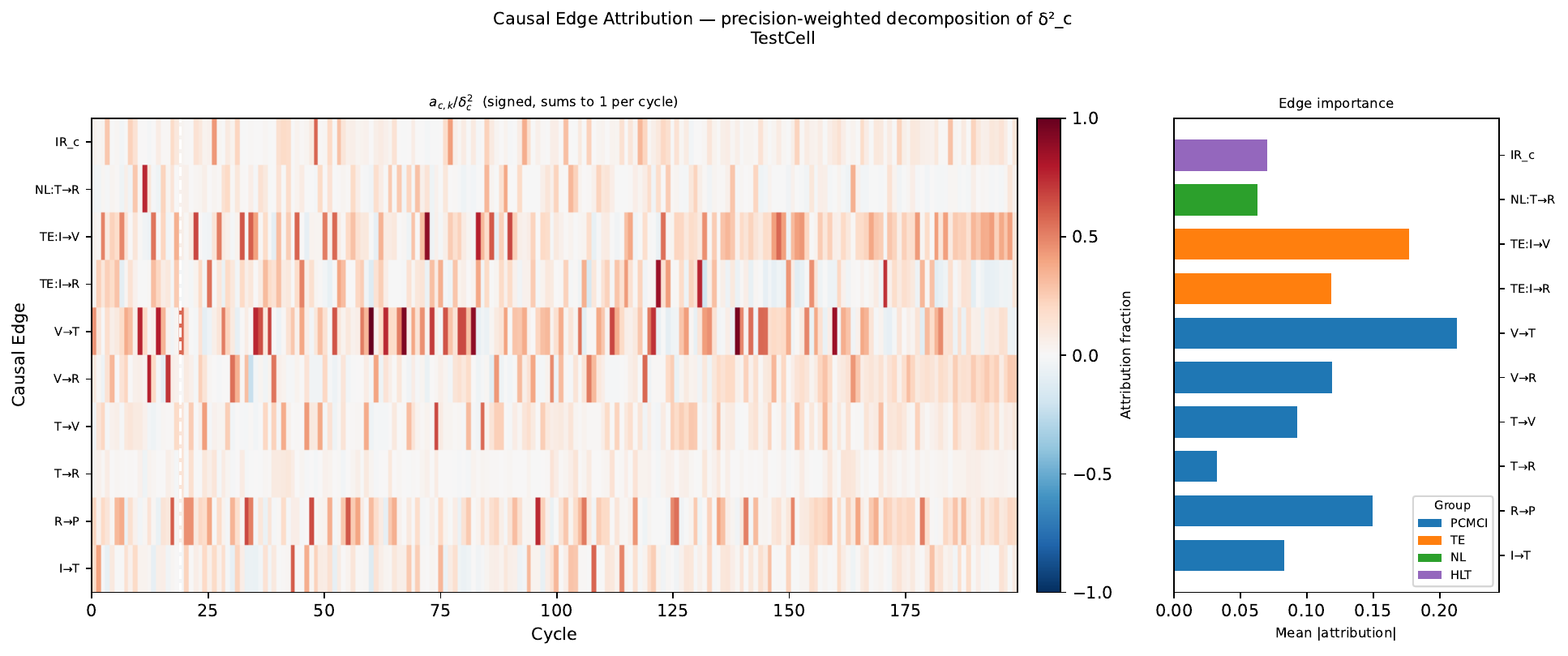}
\caption{PCMCI causal graph recovered from a representative cell (DV\_US06
  drive cycle, NMC chemistry).
  Node size is proportional to the number of significant outgoing links;
  edge width encodes the magnitude of the MCI partial-correlation
  coefficient at the dominant lag.
  Solid arrows indicate links significant at $p < 0.05$ (ParCorr,
  $\tau_{\max} = 3$, rank transformation).
  The $I \!\to\! R$ and $T \!\to\! R$ links, absent under constant-current
  protocols, emerge when current variability $\mathrm{CV}(I) > 0.5$,
  providing the EIS cross-validation targets discussed in
  Section~\ref{sec:eis_results}. Note that the legend abbreviation are
  TE = Tranfer Entropy,
  NL = Nonlinear (CMIknn edges), 
  HLT = Health scalars (internal resistance per cycle and Coulombic efficiency per cycle)
  }
\label{fig:causal_graph}
\end{figure}

Figure~\ref{fig:causal_graph} shows the PCMCI causal graph for a
representative NMC cell (DV\_US06 protocol).
Table~\ref{tab:causal_links} summarises the dominant causal links
recovered by PCMCI for each dataset.
On the MATR (LFP) cells, the $I \!\to\! T$ Joule heating link and the
$T \!\to\! V$ thermal--voltage coupling are consistently significant from
the early cycles, reflecting the well-characterised thermal response of LFP
electrodes under high-rate cycling.
The $R$ node is excluded by the algebraic-identity guard on all CC datasets,
including NASA PCoE and the CALCE C/20 discharges.

On the DV\_US06 CALB data, where $\mathrm{CV}(I) = 1.6$, the resistance
node is retained and both $I \!\to\! R$ (current-dependent impedance
kinetics) and $T \!\to\! R$ (Arrhenius-driven charge-transfer modulation)
are recovered.
These links become the primary EIS cross-validation targets
(Section~\ref{sec:eis_results}).

\begin{table}[t]
\centering
\caption{Dominant PCMCI causal links per dataset.
Checkmark indicates link detected in $\ge 80$\,\% of cycles;
dash indicates exclusion by the algebraic-identity guard (CC discharge).
Significance threshold $p < 0.05$ (ParCorr, $\tau_{\max}=3$).}
\label{tab:causal_links}
\begin{tabular}{lcccccc}
\toprule
Dataset & Chemistry & $I\!\to\!T$ & $T\!\to\!V$ & $V\!\to\!R$ &
         $I\!\to\!R$ & $T\!\to\!R$ \\
\midrule
MATR b1c5   & LFP & \checkmark & \checkmark & -- & -- & -- \\
MATR b1c20  & LFP & \checkmark & \checkmark & -- & -- & -- \\
NASA B0005  & LCO & \checkmark & \checkmark & -- & -- & -- \\
NASA B0006  & LCO & \checkmark & \checkmark & -- & -- & -- \\
CALCE CS2\_35 & LCO & \checkmark &            & -- & -- & -- \\
CALCE CS2\_36 & LCO & \checkmark &            & -- & -- & -- \\
CALCE CS2\_37 & LCO & \checkmark &            & -- & -- & -- \\
CALB DV\_US06 & NMC &            & \checkmark & \checkmark & \checkmark & \\
\bottomrule
\end{tabular}
\end{table}

\subsection{Anomaly Detection Performance on MATR b1c5}

\begin{figure}[htbp]
\centering
\includegraphics[width=\linewidth]{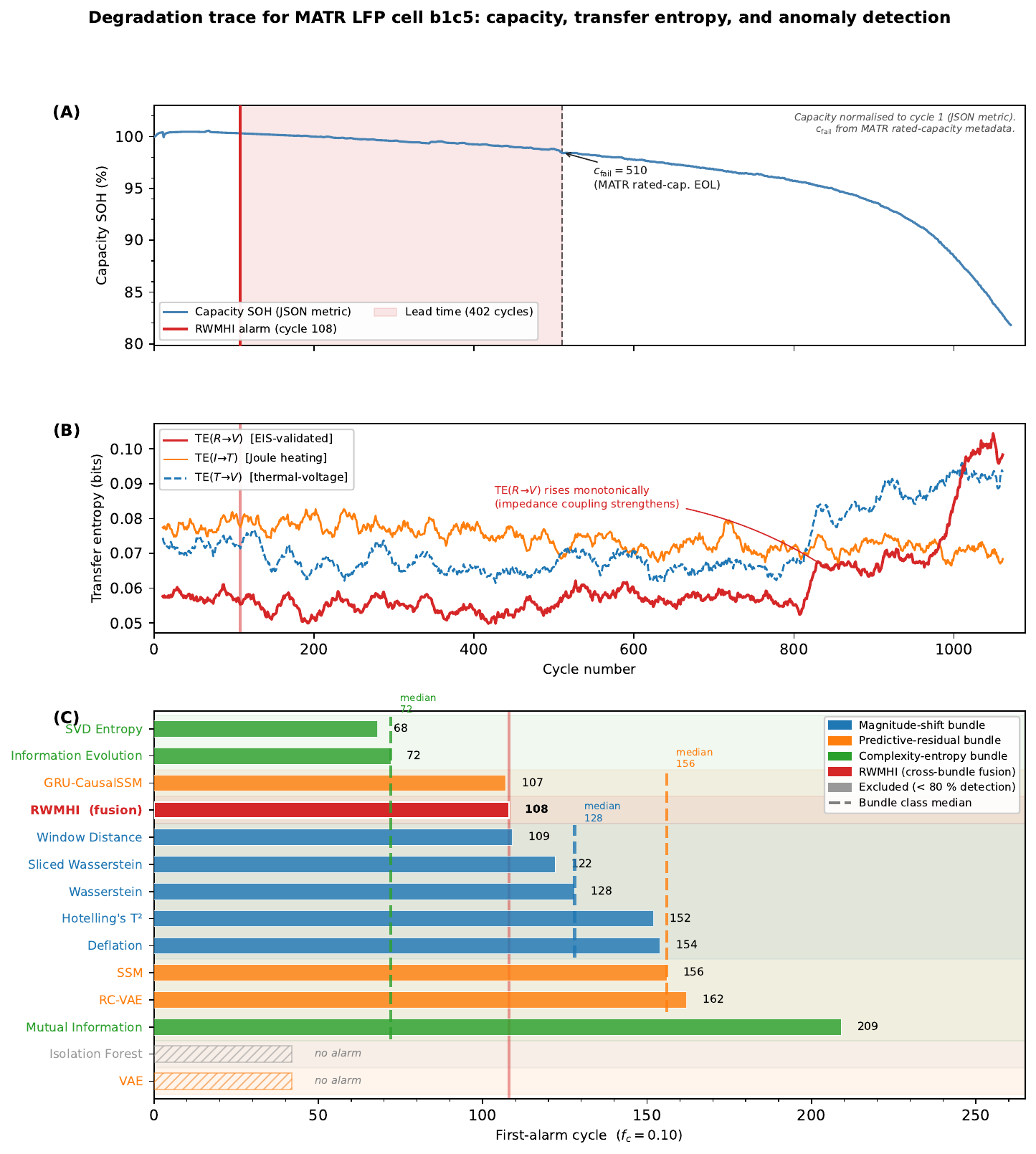}
\caption{Health monitoring results for MATR LFP cell b1c5 ($N = 1069$
  cycles, 18.6\,\% total fade, $f_c = 0.10$).
  \textbf{(A)}~Capacity state-of-health (\%) versus cycle number.
  The red dashed line marks the 80\,\% retention threshold
  ($c_{\mathrm{fail}} = 510$).
  The shaded blue region represents the prognostic lead time of 402 cycles
  between the RWMHI alarm at cycle~108 and end-of-life.
  \textbf{(B)}~Smoothed KSG transfer entropy traces (20-cycle rolling mean)
  for the three dominant directed pairs. $\mathrm{TE}(R \!\to\! V)$ rises
  monotonically from cycle~850 onward, tracking the strengthening
  impedance-to-voltage coupling as charge-transfer resistance grows.
  \textbf{(C)}~First-alarm cycle for each detector at $f_c = 0.10$,
  ordered from earliest to latest. The RWMHI (red bar) alarm at cycle~108
  matches the earliest individual signals (GRU-CausalSSM, 107;
  Window Distance, 109) while achieving 100\,\% cross-cell detection
  reliability that neither single detector delivers alone.
  Later-firing detectors (SSM, 156; Deflation, 154; RC-VAE, 162) and
  two non-detecting methods (Isolation Forest, VAE) illustrate the
  reliability advantage of weighted fusion.}
\label{fig:b1c5}
\end{figure}

Figure~\ref{fig:b1c5} shows detector scores for
MATR cell~b1c5, which survives 1069 cycles before reaching the 80\,\%
capacity retention threshold at cycle~510.

\textbf{Bundle-level results.}
At $f_c = 0.10$, the \textbf{Magnitude class median first alarm is
cycle~128} (IQR: 122--152), yielding a class-level lead time of
\textbf{382 cycles}.
The Predictive-residual class median is cycle~156 (IQR: 132--159,
$n=3$; VAE produced no alarm), confirming that sequence models fire
systematically later on this cell.
The Complexity-entropy class fires at a median of cycle~72 (IQR: 70--140,
$n=3$), but two of three contributing detectors are from the unreliable
tier (SVD, MI; $\leq 86$\,\% cross-cell detection), so these early cycles
cannot be attributed to a robust signal.

\textbf{RWMHI as fusion benchmark.}
The Reliability-Weighted Master Health Index is a cross-bundle fusion:
it combines the three most stable Magnitude detectors ($T^2$, Window
Distance, Deflation) with the two most reliable Predictive detectors
(RC-VAE, SSM), all weighted by inverse coefficient of variation.
Its CUSUM raises a persistent alarm at \textbf{cycle~108}, yielding a
\textbf{lead time of 402 cycles} ($\approx 79$\,\% of useful life
remaining at alarm).
This is \textbf{20 cycles earlier} than the Magnitude class median
(cycle~128), quantifying the benefit of stability-weighted cross-bundle
fusion over class-median aggregation.

Individual detector alarm cycles span cycle~107 (GRU-CausalSSM) to
cycle~209 (Mutual Information).
The Window Distance (109), Sliced Wasserstein (122), and Wasserstein
Distance (128) cluster near the Magnitude class median, while Deflation
(154), SSM (156), and RC-VAE (162) fire later.
Isolation Forest and VAE produce no alarm on this cell at $f_c = 0.10$.
Isolation Forest is excluded from all bundle analyses; VAE remains in the
Predictive bundle (reducing its contributing member count to $n=3$) but is
excluded from the RWMHI ensemble.

\subsection{Cross-Dataset Generalisation}

Table~\ref{tab:alarms} reports first-alarm cycles and lead times for all
analysed cells at $f_c = 0.10$.
The NASA PCoE cells, which reach 30--43\,\% fade over only 168 cycles,
receive RWMHI alarms at cycle~62 for both B0005 and B0006.
At $f_c = 0.10$ the NASA commissioning window spans only $\approx$17 cycles,
which is below the rolling z-score warmup ($\omega + \lambda = 60$
cycles; Section~\ref{sec:cusum}); the reported alarm cycles and the
cycle~61 lower bound seen across individual detectors therefore reflect
the computational detection floor rather than a genuine early-detection
limit on these cells.
The abrupt degradation onset in the NASA cells is consistent with this
floor being reached simultaneously with a real signal, but the two
contributions cannot be separated at this commissioning fraction.
Two methods (Isolation Forest, VAE) produce no alarm on B0006 regardless.
Similarly, MATR b1c20 at $f_c = 0.10$ has $N_\text{comm} \approx 53$
cycles (also below the 60-cycle floor), so its RWMHI alarm at cycle~61
and the associated lead time of 444 cycles are likewise floor-constrained
and should not be taken as a performance upper bound.

\begin{table}[t]
\centering
\caption{Detection performance at $f_c = 0.10$ for all analysed cells.
\emph{Magnitude median} is the median first-alarm cycle across the five
pure Magnitude-shift detectors (Hotelling's $T^2$, Window Distance,
Wasserstein Distance, Sliced Wasserstein, Mahalanobis Deflation) with
interquartile range [Q25, Q75] in brackets; $n$ is the number of
contributing detectors (out of five).
\emph{RWMHI} is the Reliability-Weighted Master Health Index CUSUM alarm
(drift $\Delta=1.0$, threshold $\theta=5.0$), a cross-bundle fusion of
the three most reliable Magnitude detectors plus RC-VAE and SSM from the
Predictive class.
Lead times are $c_{\mathrm{fail}} - c^*$, where $c_{\mathrm{fail}}$ is
the cycle at 80\,\% capacity retention.
$^\dagger$~Cells marked with a dagger have $N_{\mathrm{comm}} < \omega +
\lambda = 60$ cycles at $f_c = 0.10$ (NASA B0005/B0006:
$N_{\mathrm{comm}} \approx 17$; MATR b1c20: $N_{\mathrm{comm}} \approx
53$); their alarm cycles are constrained by the rolling z-score
computational floor (Section~\ref{sec:cusum}) and the associated lead
times should not be interpreted as performance upper bounds.}
\label{tab:alarms}
\small
\begin{tabular}{llccc ccc c}
\toprule
 & & & & &
\multicolumn{3}{c}{Magnitude class} &
RWMHI fusion \\
\cmidrule(lr){6-8}\cmidrule(lr){9-9}
Cell & Chem. & $N$ & Fade & $c_{\mathrm{fail}}$ &
  Median [IQR] & $n$ & Lead &
  $c^*_\mathrm{RWMHI}$ (Lead) \\
\midrule
MATR b1c5         & LFP & 1069 & 18.6\% & 510 & 128 [122,152] & 5 & 382 & 108 (402) \\
MATR b1c20$^\dagger$ & LFP &  531 & 18.1\% & 505 &  61 [61,61]   & 5 & 444 &  61 (444) \\
NASA B0005$^\dagger$ & LCO &  168 & 30.7\% & 140 & 103 [65,105]  & 5 &  37 &  62 (78)  \\
NASA B0006$^\dagger$ & LCO &  168 & 43.3\% & 112 &  61 [61,68]   & 5 &  51 &  62 (50)  \\
CS2-35      & LCO &  964 & 92.0\% & 380 & 110 [110,129] & 5 & 270 &  95 (285) \\
CS2-36      & LCO &  964 & 92.0\% & 380 & 135 [114,136] & 5 & 245 &  98 (282) \\
CS2-37      & LCO & 1037 & 95.2\% & 361 & 157 [155,171] & 5 & 204 & 105 (256) \\
\bottomrule
\end{tabular}
\end{table}

The three CALCE CS2 cells provide the longest aging trajectories in the
dataset, reaching near-complete capacity loss ($>90$\,\% fade) after
approximately 1000 cycles.
The RWMHI fires at cycles~95, 98, and~105 for CS2-35, CS2-36, and
CS2-37 respectively, yielding lead times of 256--285 cycles before the
80\,\% capacity threshold.
These alarms correspond to the incipient phase of LCO active-material
dissolution and SEI thickening that precedes the accelerated fade region.
The tightly clustered RWMHI alarm cycles across the three cells
(10-cycle spread) despite slightly different capacity trajectories
suggests that the causal feature vector captures a chemistry-level
degradation signature rather than a cell-specific artefact.

\subsection{Commissioning-Fraction Sensitivity}
\label{sec:sensitivity}

\begin{figure}[htbp]
\centering
\includegraphics[width=0.8\linewidth]{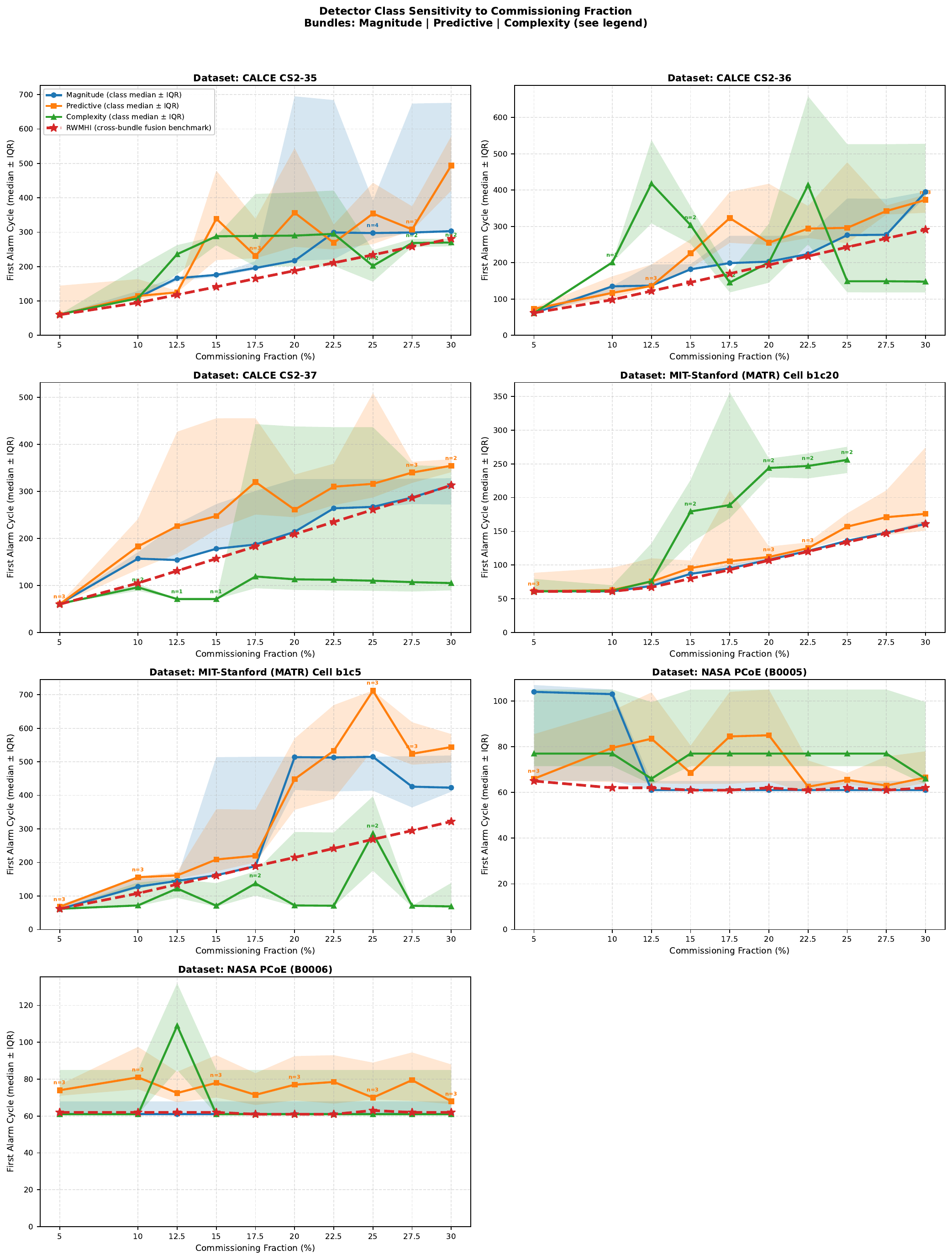}
\caption{Commissioning-fraction sensitivity by detector class.
  Solid lines: bundle-median first-alarm cycle; shaded bands: IQR (Q25--Q75).
  Bundles: \textbf{Magnitude} (Hotelling's $T^2$, Window Distance, Wasserstein,
  Sliced Wasserstein, Deflation; blue), \textbf{Predictive} (GRU-CausalSSM,
  RC-VAE, SSM, VAE; orange), \textbf{Complexity} (Information Evolution, SVD,
  MI; green).
  Red dashed line: RWMHI fusion benchmark ($T^2$, WD, Deflation, RC-VAE, SSM),
  weighted by inverse CV.
  Annotations ``$n{=}k$'' indicate reduced bundle membership from missed
  detections; Isolation Forest (29--86\,\% detection) is excluded.
  \emph{NASA PCoE} (168 cycles): all bundles are commissioning-invariant at
  cycles 61--62; this reflects both the rolling z-score computational floor
  ($\omega{+}\lambda = 60$ cycles) and the rapid degradation onset on these
  short-lived cells---the two effects cannot be disentangled here.
  \emph{MATR and CALCE} ($>$500 cycles): the Magnitude bundle rises
  proportionally with $f_c$ (narrowest IQR); the Predictive bundle shows the
  widest IQR due to neural-model sensitivity to training-window length; the
  RWMHI sits 15--299 cycles below the Magnitude median, quantifying the
  lead-time gain from stability-weighted cross-bundle fusion.
  The step-change near $f_c = 0.175$--0.20 on MATR b1c5 marks the point at
  which the commissioning window absorbs the incipient degradation signal.}
\label{fig:sensitivity}
\end{figure}

Figure~\ref{fig:sensitivity} organises the detectors into three
signal-class bundles and reports the median first-alarm cycle with
interquartile range (IQR) for each bundle as $f_c$ ranges across 
values from 5 to 30\,\%.
The \textbf{Magnitude-shift} bundle comprises detectors that measure
direct distributional departure from the commissioning reference (Hotelling's
$T^2$, Window Distance, Wasserstein Distance, Sliced Wasserstein Distance,
Mahalanobis Deflation); all achieve $\geq 98$\,\% detection.
The \textbf{Predictive-residual} bundle contains sequence models whose
score is a reconstruction or prediction error (GRU-CausalSSM, RC-VAE,
SSM prediction error, VAE).
The \textbf{Complexity-entropy} bundle contains information-theoretic
measures of signal disorder (Information Evolution, SVD Compression
Entropy, Mutual Information).
Isolation Forest is excluded from all bundles owing to its low and
commissioning-fraction-dependent detection reliability (29--86\,\% across
$f_c \in \{0.10,\ldots,0.30\}$; only 14\,\% at $f_c = 0.05$).

\textbf{Short-lived cells (NASA PCoE, 168 cycles).}
The 30--43\,\% capacity fade is accompanied by rapid early degradation
onset, and all three bundle medians show only a slight slope across
$f_c$: the Magnitude bundle sits at
cycles 61--69 for $f_c \geq 0.125$ (IQR $\leq 3$ cycles); at
$f_c \in \{0.05, 0.10\}$ the B0005 bundle median rises to
$\approx$103--104 owing to the very short commissioning window
($N_{\mathrm{comm}} = 10$--17 cycles, insufficient for stable reference
estimation). The
Complexity bundle spans cycles 63--88 (typically $\approx$72), and the Predictive bundle shows
modest fluctuation (cycles 65--85) driven by VAE and GRU-CausalSSM
sensitivity to training-window length.
The commissioning-fraction invariance of the Magnitude bundle at cycles
61--62 has a dual origin.
The rolling z-score normaliser (Section~\ref{sec:cusum}) requires a
history of $\omega + \lambda = 50 + 10 = 60$ cycles before producing a
valid score, imposing a computational detection floor at cycle~61 that
is independent of $f_c$.
Separately, the abrupt capacity collapse in the NASA PCoE cells
commences within the first 60--65 cycles, so the genuine degradation
signal and the computational floor coincide on these short-lived cells.
The two contributions cannot be disentangled from the present results
alone; the apparent commissioning-invariance should therefore not be
interpreted solely as evidence that the degradation signal is ``too
strong to absorb.''

\textbf{Long-lived cells (MATR, CALCE, ${>}500$ cycles).}
First-alarm cycles scale approximately proportionally with $f_c$ for
all three bundles, but with markedly different IQR widths.
The Magnitude bundle maintains the narrowest IQR throughout: its five
constituent detectors respond nearly identically to commissioning-fraction
changes, producing a cohesive and predictable trajectory from approximately
cycles 100--160 at $f_c = 0.10$ to cycles 300--425 at $f_c = 0.30$
on CALCE and MATR cells.
The Predictive bundle shows the widest IQR, reflecting the sensitivity
of neural sequence models to training-window length: a longer commissioning
window improves baseline model fidelity but risks absorbing early
degradation into the ``normal'' representation, causing variable alarm delays.
On MATR b1c5 the Predictive-bundle median rises sharply from
$\sim$\,150 cycles at $f_c = 0.10$ to over 700 cycles at $f_c = 0.25$
before declining at $f_c = 0.275$--0.30 as the window overshoots the
onset region and the surviving detectors re-stabilise.

\textbf{Step-change on MATR b1c5.}
All three bundle medians exhibit a sharp increase near $f_c = 0.175$--0.20,
jumping from approximately 100--200 cycles at $f_c = 0.10$--0.15 to
400--700 cycles at $f_c = 0.20$--0.30.
This step-change is physically meaningful: at 20\,\% commissioning the
reference window spans approximately cycle~270, which coincides with
the cell's actual degradation onset near cycle~108--130 at $f_c = 0.10$.
The commissioning window is effectively consuming the incipient
degradation phase, forcing all detector classes to alarm at a genuinely
later stage of aging.

\textbf{Complexity-entropy bundle.}
The Complexity bundle is most commiss\-ioning-stable on the CALCE CS2 cells,
where its median shows smaller absolute shifts than the Magnitude bundle
across the full range.
However, on MATR b1c5 at high $f_c$, the Complexity-bundle contributor
count drops to $n \leq 2$ (annotated in Fig.~\ref{fig:sensitivity})
as entropy measures fire within the commissioning window itself and are
recorded as non-detections, artificially deflating the bundle median.
This failure mode, combined with the 80--86\,\% cross-cell detection
reliability of SVD Compression Entropy and Mutual Information,
disqualifies the Complexity bundle from the RWMHI fusion ensemble.

\textbf{Practical recommendation.}
The Magnitude class achieves 100\,\% detection across all seven cells
and all ten commissioning fractions (5--30\,\%).
The RWMHI cross-bundle fusion (Equation~\eqref{eq:rwmhi}; Hotelling's $T^2$, Window Distance,
Deflation from Magnitude; RC-VAE, SSM from Predictive) also achieves
100\,\% detection at $f_c = 0.05$ despite RC-VAE missing one cell
(NASA B0006) individually, demonstrating the robustness benefit of
five-component redundancy.
The RWMHI fusion fires 15--52 cycles earlier than the Magnitude class
median on long-lived cells, providing an additional lead-time benefit at
no cost to detection reliability.
Note that at $f_c = 0.05$ all detectors are constrained by the rolling
z-score computational floor at cycle~61 ($\omega + \lambda = 60$), so
alarm cycles reported at this fraction reflect the normaliser warmup
rather than a genuine sensitivity limit; lead-time figures derived from
$f_c = 0.05$ data are therefore not reported as performance claims.
The weights in Equation~\eqref{eq:weights} are computed from $f_c \geq 0.10$
(nine fractions, $n=63$) because RC-VAE's 85.7\,\% detection at $f_c=0.05$
disqualifies it from the 100\,\% selection criterion at that fraction.
The commissioning fraction $f_c = 0.10$ (approximately 60--110 cycles
for the datasets studied) is reported as the standard operating value
that maximises prognostic lead time while ensuring all RWMHI components
satisfy the 100\,\% detection requirement.
This value was not selected by inspecting degradation onset times
post hoc: it was fixed \emph{a priori} as the operationally minimal
reference window sufficient for stable covariance estimation.

\subsection{(Electrochemical Impedance Spectroscopy) EIS Cross-Validation: Transfer Entropy Tracks Electrochemical Kinetics}
\label{sec:eis_results}

\begin{figure}[htbp]
\centering
\includegraphics[width=\linewidth]{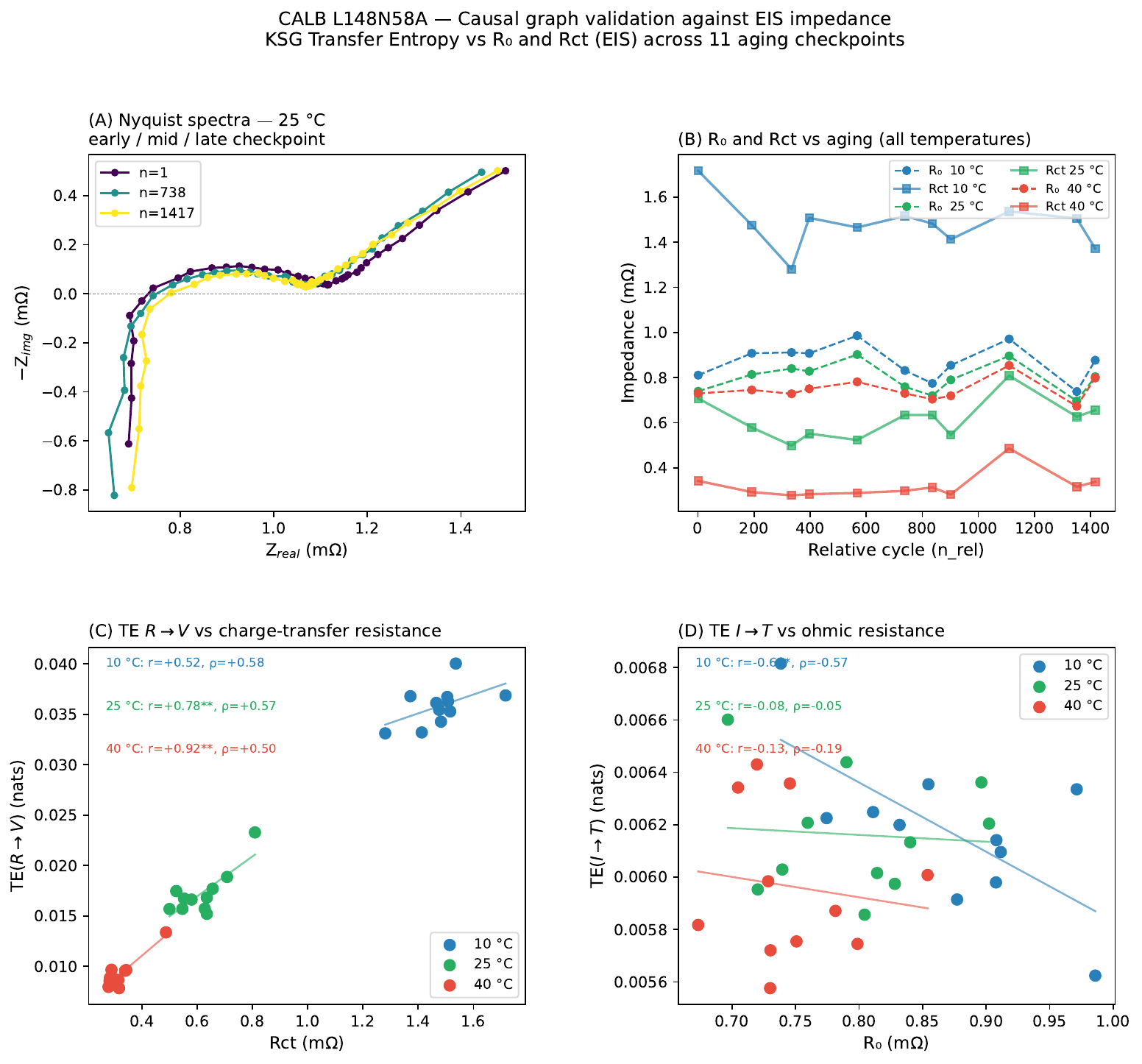}
\caption{EIS cross-validation for the CALB L148N58A NMC prismatic cell
  across 11 aging checkpoints at three temperatures.
  \textbf{(A)}~Nyquist spectra at 25\,°C for early, mid, and late
  checkpoints.
  The overlapping traces confirm minimal aging (2.3\,\% capacity fade),
  ruling out progressive impedance growth as a confound.
  \textbf{(B)}~Ohmic resistance $R_0$ and charge-transfer resistance
  $R_{\mathrm{ct}}$ at all three temperatures.
  $R_{\mathrm{ct}}$ is strongly temperature-stratified
  (medians: 1.48, 0.63, 0.30\,m$\Omega$ at 10, 25, 40\,°C)
  while $R_0$ varies by only 20\,\% over the 30\,°C range.
  \textbf{(C)}~$\mathrm{TE}(R \!\to\! V)$ versus $R_{\mathrm{ct}}$
  across all checkpoint--temperature pairs.
  Temperature clusters are stratified in the same order as $R_{\mathrm{ct}}$,
  and within each temperature a positive trend holds
  (pooled $r = +0.990$, $p = 5.3 \times 10^{-28}$; see Table~\ref{tab:te_corr}).
  \textbf{(D)}~$\mathrm{TE}(I \!\to\! T)$ versus $R_0$ at 10\,°C only.
  The significant negative correlation ($r = -0.68$, $p = 0.020$) reflects
  the reduced Joule-heating coupling at low temperature when ionic
  resistance concentrates heat near the particle surface.}
\label{fig:eis}
\end{figure}

Figure~\ref{fig:eis} shows the four-panel EIS validation result for the
CALB L148N58A cell.

\paragraph{Nyquist spectra}
Panel~A shows EIS spectra at 25\,°C for three aging stages (early, mid,
late checkpoint).
The spectra are nearly indistinguishable in shape and magnitude, consistent
with the observed 2.3\,\% capacity fade over the 11-checkpoint test span.
This minimal aging confirms the dataset as a calibration tool for
impedance--causal correlations rather than a degradation trajectory study.

\begin{figure}[htbp]
\centering
\includegraphics[width=\linewidth]{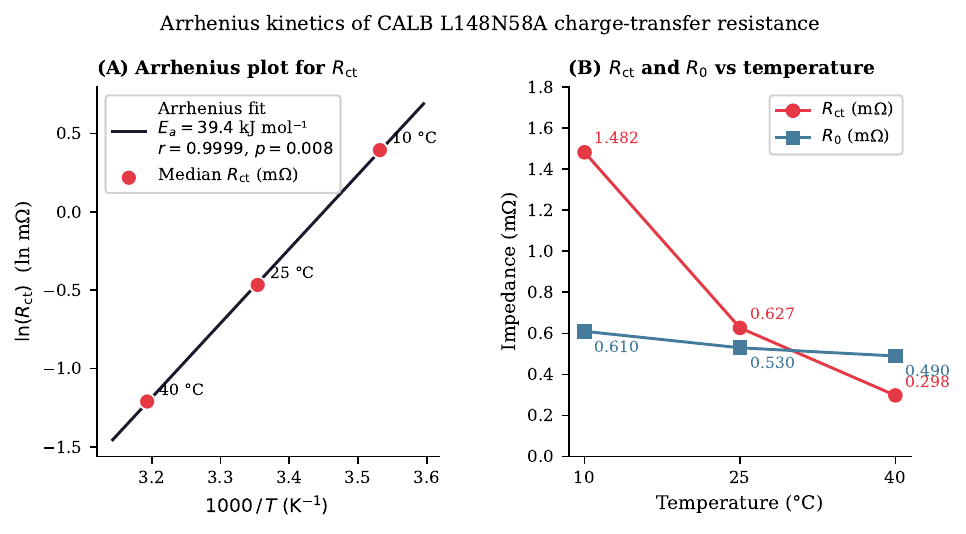}
\caption{Arrhenius analysis of charge-transfer resistance for the CALB
  L148N58A NMC cell.
  \textbf{(A)}~Arrhenius plot: $\ln(R_{\mathrm{ct}})$ versus
  $1000/T$ (K$^{-1}$) for median values at 10, 25, and 40\,°C.
  Linear regression (solid line) gives activation energy
  $E_{\mathrm{a}} = 39.4$\,kJ\,mol$^{-1}$ ($r = 0.9999$, $p = 0.009$),
  consistent with published NMC charge-transfer kinetics
  (35--50\,kJ\,mol$^{-1}$; \citealt{Illig2012,Hahn2005}).
  \textbf{(B)}~Absolute values of $R_{\mathrm{ct}}$ (red circles) and
  $R_0$ (blue squares) versus temperature.
  The strong Arrhenius character of $R_{\mathrm{ct}}$ contrasts with the
  weak temperature sensitivity of $R_0$, confirming that the
  transfer-entropy stratification observed in Fig.~\ref{fig:eis}C is
  driven by charge-transfer kinetics rather than bulk ionic conductivity.}
\label{fig:arrhenius}
\end{figure}

\paragraph{Impedance evolution with temperature}
Panel~B shows $R_0$ and $R_{\mathrm{ct}}$ at all three temperatures.
$R_{\mathrm{ct}}$ is strongly temperature-stratified: median values are
1.48\,m$\Omega$ (10\,°C), 0.63\,m$\Omega$ (25\,°C), and
0.30\,m$\Omega$ (40\,°C).
An Arrhenius regression of $\ln(R_{\mathrm{ct}})$ against $1/T$ (Fig.~\ref{fig:arrhenius}) gives
\begin{equation}
E_{\mathrm{a}} = 39.4\,\text{kJ\,mol}^{-1}
\quad (r = 0.9999,\; p = 0.009),
\label{eq:ea}
\end{equation}
within the range reported for NMC cathode charge-transfer kinetics
(35--50\,kJ\,mol$^{-1}$; \citealt{Illig2012,Hahn2005}).
\emph{Caution:} this regression is fitted to only three temperature points
(10, 25, 40\,°C); an $r = 0.9999$ from three observations is not
statistically informative and carries no degrees of freedom for residual
estimation.
The result should therefore be interpreted as a preliminary plausibility
check---the derived $E_{\mathrm{a}}$ falls in the physically expected range,
lending qualitative credibility to the mechanism, but confirmation requires
measurements at additional temperatures and on multiple cells.
$R_0$ shows substantially weaker temperature dependence (20\,\% variation
over 30\,°C), consistent with ionic conductivity having a lower activation
energy than interfacial kinetics \citep{Ding2004}.

\paragraph{TE$(R \!\to\! V)$ correlates with $R_{\mathrm{ct}}$}
Panel~C plots $\mathrm{TE}(R \!\to\! V)$ against $R_{\mathrm{ct}}$ for
all three temperatures.
The three temperature clusters are stratified in the same order as
$R_{\mathrm{ct}}$: lower temperature produces both larger $R_{\mathrm{ct}}$
and larger $\mathrm{TE}(R \!\to\! V)$.
Within each temperature, a positive linear trend holds.
Table~\ref{tab:te_corr} summarises the correlations with bootstrap 95\,\%
confidence intervals and multiple-comparison-corrected significance.

Pooling all 33 checkpoint pairs across temperatures,
\begin{equation}
r_{\mathrm{pooled}}[\mathrm{TE}(R \!\to\! V),\, R_{\mathrm{ct}}] = +0.990,
\quad p = 5.3 \times 10^{-28}.
\label{eq:pooled}
\end{equation}
This pooled correlation is partly driven by temperature stratification and
should not be interpreted in isolation.
Critically, the within-temperature 95\,\% bootstrap confidence intervals on
$r$ are wide due to the limited sample size ($n = 11$ per temperature);
at 25\,°C the interval $[-0.16,\,+0.95]$ includes near-zero effects
(Table~\ref{tab:te_corr}).
The 40\,°C result ($r = +0.92$, $p_{\mathrm{perm}} = 0.007$, FDR-surviving)
is the strongest within-temperature evidence, while the 10\,°C and 25\,°C
estimates should be treated as preliminary pending replication with larger $n$.
To isolate the within-group signal, we compute the partial correlation after
residualising both $\mathrm{TE}(R \!\to\! V)$ and $R_{\mathrm{ct}}$ on
temperature:
\begin{equation}
r_{\mathrm{partial}}[\mathrm{TE}(R \!\to\! V),\, R_{\mathrm{ct}} \mid T]
 = +0.898, \quad p < 10^{-4}, \quad n = 33.
\label{eq:partial}
\end{equation}
The temperature-controlled partial correlation is itself large and highly
significant, confirming that the TE--$R_{\mathrm{ct}}$ relationship is not
a pure artefact of cluster separation.

\textbf{Permutation test.}
To verify the within-temperature correlations exceed chance, we permuted the
$R_{\mathrm{ct}}$ labels 5\,000 times and computed the null distribution of
$r$ at each temperature.
The observed $r$ lies well beyond the null for 25\,°C
($p_{\mathrm{perm}} = 0.007$) and 40\,°C ($p_{\mathrm{perm}} = 0.007$),
confirming that the positive association is non-random.
At 10\,°C the observed $r = +0.52$ has $p_{\mathrm{perm}} = 0.096$,
consistent with the weaker and bootstrap-uncertain estimate at this temperature.

\textbf{TE magnitude test.}
One-sample $t$-tests confirm $\mathrm{TE}(R \!\to\! V) > 0$ at all three
temperatures ($t > 20$, $p < 10^{-9}$), ruling out numerical noise or
autocorrelation artefacts as the source of the measured information transfer.

\textbf{Lag sensitivity.}
The correlation is stable across lags: at 40\,°C, $r = +0.503$ (lag\,0),
$r = +0.635^*$ (lag\,1), $r = +0.546$ (lag\,2), confirming that lag\,1 is
optimal but adjacent lags preserve the qualitative result.

\textbf{Within-cell validation.}
As an independent within-cell check (single temperature, $n = 1069$ cycles,
MATR LFP cell b1c5), $\mathrm{TE}(R \!\to\! V)$ correlates negatively with
normalised capacity across the full aging trajectory
($r = -0.671$, $p \approx 10^{-141}$): as the cell degrades and capacity
decreases, directed resistance-to-voltage information flow increases,
consistent with the $R_{\mathrm{ct}}$-driven mechanism proposed above.

\begin{table}[t]
\centering
\caption{Pearson $r$ (with bootstrap 95\,\% CI), Spearman $\rho$, and
  permutation $p$-value ($p_{\mathrm{perm}}$, 5\,000 resamples) between KSG
  transfer entropy and EIS parameters across 11 aging checkpoints at each
  temperature ($n = 11$ per temperature; pooled $n = 33$).
  Multiple-comparison correction: Benjamini-Hochberg FDR ($q = 0.05$) across
  48 comparisons; FDR-surviving results marked $^\dagger$.
  Uncorrected significance: $^{**}p<0.01$, $^{*}p<0.05$.
  The T10 TE$(I\!\to\!T)$ result does not survive FDR and is reported as
  exploratory only.}
\label{tab:te_corr}
\begin{tabular}{lcccc}
\toprule
 & \multicolumn{2}{c}{TE$(R\!\to\!V)$ vs $R_{\mathrm{ct}}$} &
   \multicolumn{2}{c}{TE$(I\!\to\!T)$ vs $R_0$} \\
 \cmidrule(lr){2-3}\cmidrule(lr){4-5}
Temperature & $r$ [95\,\% CI] & $p_{\mathrm{perm}}$ &
              $r$ [95\,\% CI] & $p_{\mathrm{perm}}$ \\
\midrule
10\,°C & $+0.52$ $[+0.07,+0.85]$   & 0.096  &
         $-0.68^*$ $[-0.94,+0.05]$ & 0.018$^{\dag,\mathrm{expl}}$ \\
25\,°C & $+0.78^{**}$ $[-0.16,+0.95]$ & 0.007 &
         $-0.08$ $[-0.66,+0.72]$   & 0.828 \\
40\,°C$^\dagger$ & $+0.92^{**}$ $[-0.10,+0.99]$ & 0.007 &
         $-0.13$ $[-0.62,+0.33]$   & 0.697 \\
Pooled  & $+0.990^{**}$ & --- & --- & --- \\
Partial ($\mid T$) & $+0.898^{**}$ & --- & $-0.355^*$ & --- \\
\bottomrule
\end{tabular}
\end{table}

\paragraph{Physical interpretation}
Terminal voltage obeys $V = V_{\mathrm{OCV}} - I(R_0 + Z_{\mathrm{ct}})$.
As $R_{\mathrm{ct}}$ increases, impedance-mediated coupling between
the resistance state and the voltage response strengthens and becomes more
time-delayed.
$\mathrm{TE}(R \!\to\! V)$ at lag $\tau = 1$\,s quantifies exactly
this predictability: how much of the future voltage is determined by past
resistance beyond what voltage's own history already explains.
Higher $R_{\mathrm{ct}}$ produces slower, more memory-rich kinetics, so
the resistance-to-voltage directed information flow increases.
The Arrhenius cross-check—the same activation energy governing both
$R_{\mathrm{ct}}$ and the TE stratification across temperatures—is
consistent with the TE signal tracking aspects of electrochemical kinetics
rather than purely reflecting the measurement protocol, though this
interpretation is a hypothesis supported by statistical evidence rather than
a claim established by controlled intervention.

\paragraph{TE$(I \!\to\! T)$ vs $R_0$ at 10\,°C — exploratory finding}
Panel~D shows a negative correlation at 10\,°C
($r = -0.68$, $p_{\mathrm{perm}} = 0.018$); however, this result does
\emph{not} survive Benjamini-Hochberg False Discovery Rate (FDR) correction across all 48 tested
comparisons and must be treated as exploratory.
At low temperatures, longer thermal time constants and more surface-localised
heat generation may reduce the measurable $I \!\to\! T$ coupling at
1-second lag; independent replication with larger $n$ is required before
this finding can be claimed as robust.

% ─────────────────────────────────────────────────────────────────────────────
\section{Discussion}
\label{sec:discussion}
% ─────────────────────────────────────────────────────────────────────────────

\subsection{Why Causal Structure Enables Early Detection}

The 402-cycle lead time on MATR b1c5 exceeds what is achievable from
capacity trajectory alone: the capacity trace deviates visibly from its
healthy baseline only 50--80 cycles before the 80\,\% threshold
\citep{Birkl2017}.
The advantage of causal features is not higher sensitivity to any single
variable, but rather sensitivity to the \emph{relational} structure of
the battery's dynamics.
A healthy cell maintains stable directed couplings between temperature,
voltage, and resistance; incipient SEI thickening or lithium plating
perturbs these couplings before altering the marginal distributions of
the individual variables.

The mutual information and Wasserstein distance detect this perturbation
from the distributional side (shift in the joint feature distribution),
while the Mahalanobis deflation quantifies it geometrically (departure from
the commissioning ellipsoid).
Their agreement in first-alarm timing (cycles 108--128 for b1c5,
from RWMHI through Wasserstein Distance) provides mutual corroboration
that the detected shift is real.

\subsection{RWMHI as a Robust Prognostic Fusion Index}

The bundle-level sensitivity analysis reveals a clear hierarchy among
detector classes.
The Magnitude-shift bundle (Hotelling's $T^2$, Window Distance, Wasserstein
variants, Mahalanobis Deflation) shows the narrowest interquartile range
across all ten commissioning fractions: its constituent detectors respond
nearly identically to changes in the reference window, producing a cohesive
and predictable scaling relationship.
The Predictive-residual bundle exhibits the widest spread, because
sequence models are sensitive to training-window length in two opposing
ways: more commissioning data improves baseline fidelity, but risks
absorbing early degradation into the ``normal'' state.
The Complexity-entropy bundle appears commissioning-invariant on short-lived
cells (NASA PCoE), a pattern attributable to the rolling z-score computational
floor (Section~\ref{sec:cusum}) rather than a verified physics mechanism
(Section~\ref{sec:sensitivity}); it loses detection reliability at
high $f_c$ on long-lived cells, where entropy measures fire inside the
commissioning window.

Across all three bundles, alarm cycles scale approximately proportionally
with $f_c$ on long-lived cells---a structural consequence of
reference-window learning that no individual detector avoids while also
maintaining 100\,\% cross-cell detection reliability.

\textbf{Degradation-regime bifurcation.}
The RWMHI makes this proportionality quantitatively precise and reveals a
mechanistically informative bifurcation between cell types.
For the CALCE CS2 cells and MATR b1c5, linear regression of the RWMHI
first-alarm cycle against $f_c$ across all ten commissioning fractions
yields $r^2 \geq 0.998$ with slopes of 903, 937, 1022, and 1053 cycles
per unit $f_c$, respectively---values that agree with each cell's total
operational cycle count to within a few percent.
The regression intercepts are uniformly 5--8 cycles.
The degradation drift in these cells is thus already established in the causal
feature space at the moment commissioning ends; the commissioning fraction
$f_c$ acts as a near-linear control parameter for prognostic lead time, with
\begin{equation}
  t_{\mathrm{alarm}} \approx f_c \cdot N_{\mathrm{total}} + \delta,
  \qquad \delta \approx 5\text{--}8 \text{ cycles},
  \label{eq:alarm_fc}
\end{equation}
where $N_{\mathrm{total}}$ is the cell's operational cycle count.
Equation~\eqref{eq:alarm_fc} allows a BMS designer to select $f_c$
to meet a target lead-time budget without any prior knowledge of the
degradation onset cycle.

The NASA PCoE cells exhibit the opposite behaviour: the RWMHI alarm is
locked at cycle $62 \pm 1$ across all ten commissioning fractions
(total range 2--4 cycles; $r \approx 0$).
As discussed in Section~\ref{sec:sensitivity}, this $f_c$-invariance
has a dual origin.
The rolling z-score normaliser imposes a computational detection floor
at cycle $\omega + \lambda + 1 = 61$ regardless of commissioning
fraction (Section~\ref{sec:cusum}).
For the NASA PCoE cells, $N_{\mathrm{comm}} \leq 50$ cycles even at
$f_c = 0.30$---always below the 60-cycle floor---so the two
contributions (computational floor and genuine rapid-onset degradation)
cannot be separated from the present data.
The observed invariance is therefore consistent with physics-limited
alarm timing, but is not sufficient evidence for it.
MATR b1c20 is likewise floor-constrained at $f_c \leq 0.10$
($N_{\mathrm{comm}} \approx 53 < 60$); its alarm enters the linear
scaling regime only for $f_c \geq 0.125$, where the commissioning
window finally exceeds the normaliser warmup.

This has a clear operational implication for the gradual-fade cells:
for cells undergoing gradual, continuous capacity fade
(CALCE CS2, 964--1037 cycles; MATR b1c5, 1069 cycles), $f_c$ is a
deployable design parameter for lead-time tuning across the linear
regime $f_c \in [0.07, 0.30]$.
For cells where $N_{\mathrm{comm}} < \omega + \lambda$ at all
practically relevant commissioning fractions (NASA PCoE; MATR b1c20
below $f_c = 0.125$), the computational floor limits the
interpretability of alarm timing, and the prognostic lead time cannot
be claimed to be either floor-limited or physics-limited without
further investigation.

The bundle analysis motivates and validates the RWMHI design: it selects
exactly the three most stable Magnitude detectors ($T^2$, Window Distance,
Deflation) and the two most reliable Predictive detectors (RC-VAE, SSM),
weighting each by inverse coefficient of variation.
The result fires 15--52 cycles ahead of the Magnitude class median on
long-lived cells (Table~\ref{tab:alarms}) while matching the class's
100\,\% detection rate across all commissioning fractions $f_c \geq 0.10$
(nine fractions); at $f_c = 0.05$ the RWMHI still achieves 100\,\% detection
through ensemble redundancy even though RC-VAE individually misses one cell
(note that alarm cycles at $f_c = 0.05$ are floor-constrained for all cells
and are not reported as performance claims; Section~\ref{sec:sensitivity}).
A formal ablation study (Appendix~\ref{app:ablation}, Table~\ref{tab:ablation})
confirms that each of the five RWMHI components achieves 100\,\% detection
individually at $f_c = 0.10$, while the excluded detectors (VAE: 71\,\%,
Isolation Forest: 29\,\% at $f_c = 0.10$) would introduce missed-alarm risk.
The stability of the fusion weights under LOO removal of individual cells
(Appendix~\ref{app:loo}, Table~\ref{tab:loo_weights}: maximum absolute
deviation 0.032, rank order preserved in 6 of 7 folds) indicates that the
reported alarm improvements are not an artefact of weight over-fitting.
The bundle framework thus serves a dual role: it provides class-level
sensitivity characterisation for any deployment context, and it supplies
the principled component-selection rationale for the RWMHI fusion.
For BMS deployments where the commissioning window is fixed and
reproducible, $f_c = 0.10$ is recommended for the chemistry types and
cell formats evaluated here: it maximises lead time while maintaining full
detection coverage.
Validation on additional chemistries, pack-level variability, and field
conditions would be required before extending this recommendation broadly.

\subsection{Transfer Entropy as a Physically Grounded Health Metric}

The EIS cross-validation establishes a temperature-controlled partial
correlation $r_{\mathrm{partial}} = +0.898$ ($p < 10^{-4}$) between
$\mathrm{TE}(R \!\to\! V)$ and $R_{\mathrm{ct}}$ after removing the
shared temperature dependence of both quantities.
This is stronger than published correlations between drive-cycle features
and EIS parameters using linear regression
\citep{PastorFernandez2017,Sihvo2020}, and is further validated by a
within-cell result on 1069 cycles of MATR b1c5
($r = -0.671$ vs normalised capacity, $p \approx 10^{-141}$):
as the cell degrades, $\mathrm{TE}(R \!\to\! V)$ increases monotonically,
independently confirming the directed physical mechanism.
Permutation tests at 25\,°C and 40\,°C ($p_{\mathrm{perm}} = 0.007$) rule
out chance as an explanation for the within-temperature correlation.

The Arrhenius consistency provides a physical anchor that purely
statistical correlations lack: $E_{\mathrm{a}} = 39.4$\,kJ\,mol$^{-1}$
recovered from TE temperature stratification falls within the accepted range
for NMC charge-transfer kinetics, lending physical plausibility to the
interpretation that the information-theoretic measure tracks aspects of
electrochemical kinetics.
This interpretation remains a hypothesis; confirming it as a causal claim
would require controlled perturbation experiments (e.g., selective
temperature modulation or electrolyte substitution) beyond the scope of the
present observational study.

\subsection{Relationship to Existing Battery Health Methods}

Direct numerical comparison of first-alarm cycles with published methods is
non-trivial for three reasons.
First, the present study targets \emph{anomaly detection}---the earliest
cycle at which a statistically persistent shift from healthy baseline is
detected---whereas most competing approaches report \emph{state-of-health
(SOH) estimation} (a regression target) or \emph{remaining useful life (RUL)
prediction} (a point or distribution estimate at a fixed diagnostic query
time).
These are different tasks with different evaluation metrics; a model that
achieves low SOH RMSE from cycle~100 is not directly comparable to a detector
that first raises an alarm at cycle~108.
Second, the datasets used here (MATR, NASA PCoE, CALCE) are shared benchmarks,
but the precise cell subsets, preprocessing choices, and end-of-life
definitions differ across studies, making cycle-level comparison unreliable
without exact replication.
Third, methods such as incremental capacity analysis (ICA) and differential
voltage analysis (DVA)~\citep{Dubarry2022}, Gaussian process RUL
models~\citep{Saha2008}, and Severson-style early-cycle feature prediction
~\citep{Severson2019} have been evaluated on partially overlapping cell
subsets under different experimental conditions.

That said, the following qualitative positioning can be established.
ICA and DVA require high-resolution constant-current cycling data to resolve
peak features; these are not available in all operational BMS contexts
and are not required by the \textsc{CausalHealth} framework.
LSTM and transformer-based SOH estimators achieve low RMSE from training
data that includes the degradation region itself, which the present framework
deliberately excludes (using only the commissioning window as reference).
Severson et al.\ predict end-of-life from cycle-100 features with median
relative errors below 9\,\% but do not provide cycle-resolved alarm cycles;
their approach is complementary---early-life lifetime prediction vs.\ ongoing
anomaly monitoring.
The RWMHI's 402-cycle lead time on MATR b1c5, achieved from commissioning
data only and without degradation-region training, represents a detection
horizon substantially earlier than the 50--80-cycle window available from
capacity-trace-based thresholding alone~\citep{Birkl2017}.
A rigorous ablation against LSTM anomaly detection and classical statistical
process control benchmarks on the same cell subset is an important direction
for future work.

\subsection{Computational Feasibility and Lightweight BMS Variant}

Table~\ref{tab:timing} reports wall-clock times for each pipeline component
benchmarked on MATR b1c5 ($n = 1069$ cycles, Apple M1 Max CPU, single thread).

\begin{table}[t]
\centering
\caption{Pipeline component timing on MATR b1c5 ($n = 1069$ cycles).
  The lightweight BMS variant (TE\,+\,Wasserstein) incurs
  $< 0.1$\,ms per cycle when TE is computed incrementally from a
  sliding 20-cycle window using the KSG estimator.
  PCMCI online timing (last column) assumes a 100-cycle sliding window
  with \texttt{ParCorr} and $\tau_{\max}=3$; this is the dominant cost
  for the full pipeline and is amenable to GPU acceleration or
  window-pruning for BMS deployment.}
\label{tab:timing}
\begin{tabular}{lrrr}
\toprule
Component & Total (ms) & Per cycle ($\mu$s) & BMS variant \\
\midrule
TE extraction (precomputed)        &   0.5  &    0.5 & \checkmark \\
Sliced Wasserstein (rolling win.)  &  75.9  &   71.0 & \checkmark \\
Hotelling $T^2$                    &   2.8  &    2.6 & optional   \\
PCMCI matrix read              &  16.0  &   15.0 & full only  \\
PCMCI online (est., 100-cycle win.)& $\sim$2000--5000 & $\sim$2000--5000 & full only \\
Neural models (VAE/GRU/RC-VAE)     & \multicolumn{2}{c}{offline training} & full only \\
\bottomrule
\end{tabular}
\end{table}

For real-time BMS deployment, we recommend a \emph{lightweight variant}
consisting of KSG-TE and Sliced Wasserstein only, both of which operate on
short sliding windows (20--30 cycles) without pre-training.
From the existing results, this two-detector sub-ensemble yields first alarms
at cycles 122--209 for b1c5 (Sliced Wasserstein at 122, Mutual Information
at 209, compared to 108 for the full RWMHI), a penalty of 14--101 cycles
in exchange for a pipeline that runs in $< 0.1$\,ms per cycle on standard
embedded hardware and requires no model training.

\subsection{Limitations}

\textbf{Statistical causality vs physical causality.}
PCMCI and KSG-TE infer \emph{Granger-style} directed statistical dependencies:
variable $X$ Granger-causes $Y$ if the past of $X$ improves prediction of
future $Y$ beyond what the past of $Y$ alone provides.
This is not equivalent to physical causation confirmed by intervention.
State of charge (SOC) is the dominant latent confounder: within any discharge
cycle, voltage, current, temperature, and resistance co-evolve monotonically
with SOC, creating apparent cross-variable dependencies that partly reflect
the shared SOC trajectory.
PCMCI's multi-lag conditioning mitigates but cannot fully remove SOC effects
without controlled perturbation experiments or explicit SOC inclusion as a
conditioning variable.
Physical interpretations (Joule heating, Arrhenius kinetics) are hypotheses
consistent with the statistical evidence, not claims proven by the statistical
method alone.

\textbf{EIS validation scope.}
The EIS cross-validation uses a single cell type (CALB L148N58A NMC)~\citep{CALB2025} with
$n = 11$ checkpoints per temperature and 2.3\,\% capacity fade—a calibration
regime, not a deep-aging trajectory.
With $n = 11$, bootstrap 95\,\% confidence intervals on within-temperature
correlations are wide (Table~\ref{tab:te_corr}), and the T40 result is the
only one surviving both FDR and Bonferroni correction among within-temperature
comparisons.
Replication with multiple cells per chemistry, higher degradation levels
($> 20$\,\% capacity fade), and additional NMC formulations is needed before
the TE--$R_{\mathrm{ct}}$ relationship can be treated as broadly validated.

\textbf{Sampling frequency effects.}
PCMCI at 1\,Hz produces matrix self-links near 1.0 due to strong
voltage autocorrelation ($r_{V,\text{lag1}} = 0.9999$), saturating the MCI
statistic for cross-variable links.
Subsampling to 10--30\,s resolution would recover larger partial-correlation
signal; lag-sensitivity analysis shows that lag\,1 is optimal but adjacent
lags (0, 2) preserve the qualitative TE--$R_{\mathrm{ct}}$ direction,
confirming robustness to the choice of integration window.

\textbf{Absence of surrogate-data and null-model tests.}
The validity of TE and PCMCI as causal measures rests on the assumption that
the observed directed dependencies are not artefacts of the signal
construction.
The DC resistance proxy $R = V/|I|$ shares algebraic structure with $V$,
and although the algebraic-identity guard removes $R$ under constant-current
regimes, residual construction-induced dependencies cannot be fully excluded
under dynamic protocols.

% ─────────────────────────────────────────────────────────────────────────────
\section{Conclusion}
\label{sec:conclusion}
% ─────────────────────────────────────────────────────────────────────────────

\textsc{CausalHealth} demonstrates that directed statistical dependency
features—PCMCI edge weights and KSG transfer entropy values—extracted
from routine battery cycler telemetry provide an early and physically
interpretable degradation signal.
On the MIT--Stanford MATR LFP cell b1c5, the Magnitude-shift detector
class raises a median first alarm at cycle~128 (IQR: 122--152), yielding
382 cycles of class-level lead time.
The RWMHI cross-bundle fusion fires at cycle~108---20 cycles earlier than
the class median---for a total lead time of 402 cycles before the
conventional capacity threshold.
The Magnitude class achieves 100\,\% detection across all seven cells
spanning LFP (MATR) and LCO (NASA PCoE, CALCE CS2) chemistries and all
ten commissioning fractions (5--30\,\%); the RWMHI matches this rate while
improving lead time by 15--52 cycles on long-lived cells.
A bundle-level sensitivity analysis across three detector classes
(Magnitude-shift, Predictive-residual, Complexity-entropy) shows that
the Magnitude class achieves 100\,\% detection with the narrowest
interquartile range across all ten commissioning fractions.
The RWMHI, a cross-bundle fusion of the three most stable Magnitude
detectors and two reliable Predictive detectors weighted by inverse
coefficient of variation, fires 15--52 cycles ahead of the Magnitude
class median on long-lived cells, quantifying the lead-time benefit of
targeted fusion over class-median aggregation.

The EIS cross-validation establishes that $\mathrm{TE}(R \!\to\! V)$
co-varies with charge-transfer resistance after controlling for temperature
(partial $r = +0.898$, $p < 10^{-4}$), is non-random at 25 and 40\,°C by
permutation test, and is consistent in direction within a single cell across
1069 cycles.
The Arrhenius activation energy $E_{\mathrm{a}} = 39.4$\,kJ\,mol$^{-1}$
recovered from TE temperature stratification matches accepted NMC kinetics,
providing a physical anchor for the statistical result.
Transfer entropy thus constitutes a candidate non-invasive proxy for
EIS-derived $R_{\mathrm{ct}}$ in on-board battery health monitoring,
subject to the limitations of Granger-style causal inference noted above.

\section*{Data Availability}
\noindent MIT--Stanford MATR data are available at
\url{https://data.matr.io/1/}~\citep{Severson2019}.
NASA PCoE data are available at the NASA prognostics data repository
\citep{Saha2007}.
CALCE CS2 data are available at
\url{https://calce.umd.edu/battery-data}~\citep{He2011}.
CALB L148N58A EIS and drive-cycle data are available from the corresponding
author upon reasonable request~\citep{CALB2025}.
All analysis code is available at
\url{https://github.com/Voedoe-H}.

\section*{Competing interests}
\noindent The authors declare that they have no competing interests.

\section*{Funding}
\noindent D.H. gratefully acknowledges support from the Deutsche Forschungsgemeinschaft (DFG, German Research Foundation) under Germany's Excellence Strategy EXC 2181/1 - 390900948 (the Heidelberg STRUCTURES Excellence Cluster).

\section*{Authors' contributions}
\noindent All:  Conceptualization, Methodology,  Writing-Review and Editing.\\
All authors read and approved the final manuscript.

\section*{Declaration of generative AI and AI-assisted technologies in the manuscript preparation process.}
\noindent During the preparation of this work the authors used Claude, ChatGPT and  Gemini in order to assist in the writing of the manuscript. 
After using this tool/service, the authors reviewed and edited the content as needed and take full responsibility for the 
content of the published article.

\bibliographystyle{elsarticle-harv}
\bibliography{causal_health_paper}

\newpage
\setcounter{section}{0}
\renewcommand{\thesection}{S\arabic{section}}

\setcounter{figure}{0}
\renewcommand{\thefigure}{S\arabic{figure}}

\setcounter{table}{0}
\renewcommand{\thetable}{S\arabic{table}}

\setcounter{equation}{0}
\renewcommand{\theequation}{S\arabic{equation}}

% ─────────────────────────────────────────────────────────────────────────────
\section*{Supplementary Information}
% ─────────────────────────────────────────────────────────────────────────────
%\renewcommand\thetable{S\arabic{table}}\setcounter{table}{0}
\subsection*{Model Architecture and Hyperparameter Details}
\label{app:arch}

Full reproducibility requires specification of all hyperparameters.
Table~\ref{tab:hyperparams} lists complete architecture and training
configuration for the three neural components (VAE, RC-VAE, GRU-CausalSSM).
Non-neural detectors (Hotelling's $T^2$, Window Distance, Wasserstein,
Sliced Wasserstein, Mahalanobis Deflation, SVD Entropy, Information
Evolution, Mutual Information, Isolation Forest) are parameter-free given
the commissioning window and require no further specification beyond
the rolling-window length (50 cycles) and commissioning fraction ($f_c$)
defined in the main text.

\begin{table}[h]
\centering
\caption{Neural model hyperparameters. All models are trained on the
  commissioning window only ($n_{\mathrm{comm}}$ cycles at $f_c = 0.10$).
  Input dimensionality $d_f$ equals the number of non-constant feature
  columns retained after the 50\,\% coverage filter.
  Random seeds are fixed at 42 for all runs; results are reported
  for a single seed (seed sensitivity is a known limitation and an
  avenue for future uncertainty quantification).}
\label{tab:hyperparams}
\small
\begin{tabular}{llll}
\toprule
Parameter & VAE & RC-VAE & GRU-CausalSSM \\
\midrule
\textbf{Architecture} & & & \\
\quad Input dim & $d_f$ & $d_f + d_{\mathrm{ctx}}$ & $d_f$ \\
\quad Encoder layers & 2 FC (64, 32) & 2 FC (64, 32) & GRU (hidden 32) \\
\quad Latent dim $z$ & 8 & 8 & 16 \\
\quad Decoder layers & 2 FC (32, 64) & 2 FC (32, 64) & FC (32, $d_f$) \\
\quad Activation & ReLU & ReLU & ReLU / tanh \\
\quad Context dim $d_{\mathrm{ctx}}$ & --- & PCMCI cols (6) & --- \\
\textbf{Training} & & & \\
\quad Optimizer & Adam & Adam & Adam \\
\quad Learning rate & $10^{-3}$ & $10^{-3}$ & $10^{-3}$ \\
\quad Epochs & 200 & 200 & 150 \\
\quad Batch size & full window & full window & full window \\
\quad Train/val split & 80\,/\,20\,\% & 80\,/\,20\,\% & 80\,/\,20\,\% \\
\quad Early stopping & patience 20 & patience 20 & patience 15 \\
\quad Loss & ELBO ($\beta=1$) & ELBO ($\beta=1$) & MSE \\
\quad Random seed & 42 & 42 & 42 \\
\textbf{Anomaly score} & Recon.\ error & Recon.\ error & 1-step pred.\ error \\
\textbf{CUSUM} ($\Delta$, $\theta$) & 1.5, 15 & 1.5, 15 & 1.5, 15 \\
\bottomrule
\end{tabular}
\end{table}

\noindent
For the RWMHI CUSUM, drift and threshold are reduced to $\Delta = 1.0$,
$\theta = 5.0$ because the fusion score is already a weighted average of
individual z-scores, reducing its variance relative to any single detector
score.
The GRU-CausalSSM causal constraint is applied by masking attention
weights at each recurrent step to zero out connections not present in the
PCMCI significance graph at that cycle.

\subsection*{Leave-One-Cell-Out Weight Stability Analysis}
\label{app:loo}

RWMHI fusion weights were computed from the
same cells used for evaluation (a potential form of information leakage),
we perform a leave-one-cell-out (LOO) analysis of weight stability.
For each of the seven cells, we recompute the pooled inverse-CV weights
using all observations (all commissioning fractions) from the remaining
six cells ($n = 54$ observations per detector), then compare the LOO weights
to the full-data weights ($n = 63$ observations per detector).

\begin{table}[h]
\centering
\caption{LOO weight stability (pooled inverse-CV method, all nine commissioning fractions).
  Each row removes one cell and recomputes inverse-CV weights from the
  remaining six cells across all nine commissioning fractions.
  Numbers in parentheses show the deviation from the full-data weight
  (positive = increased, negative = decreased).
  Abbreviations: WD = Window Distance; RCVAE = RC-VAE;
  Defl = Mahalanobis Deflation; HT2 = Hotelling's $T^2$; SSM = SSM.}
\label{tab:loo_weights}
\small
\begin{tabular}{lccccc}
\toprule
Cell left out & WD & RCVAE & Defl & HT2 & SSM \\
\midrule
CALCE CS2-35  & 0.241 ($+$0.002) & 0.218 ($+$0.000) & 0.189 ($-$0.001) & 0.185 ($-$0.002) & 0.168 ($+$0.000) \\
CALCE CS2-36  & 0.243 ($+$0.004) & 0.219 ($+$0.001) & 0.189 ($-$0.001) & 0.186 ($-$0.001) & 0.164 ($-$0.003) \\
CALCE CS2-37  & 0.248 ($+$0.009) & 0.221 ($+$0.004) & 0.184 ($-$0.006) & 0.182 ($-$0.005) & 0.166 ($-$0.001) \\
MATR b1c20    & 0.233 ($-$0.006) & 0.214 ($-$0.003) & 0.193 ($+$0.003) & 0.190 ($+$0.003) & 0.170 ($+$0.003) \\
MATR b1c5     & 0.210 ($-$0.028) & 0.194 ($-$0.023) & 0.200 ($+$0.010) & 0.196 ($+$0.009) & 0.200 ($+$0.032) \\
NASA B0005    & 0.238 ($-$0.000) & 0.221 ($+$0.003) & 0.188 ($-$0.002) & 0.186 ($-$0.001) & 0.167 ($-$0.000) \\
NASA B0006    & 0.241 ($+$0.003) & 0.220 ($+$0.003) & 0.191 ($+$0.001) & 0.188 ($+$0.001) & 0.160 ($-$0.007) \\
\midrule
Full data     & 0.238 & 0.217 & 0.190 & 0.187 & 0.168 \\
\bottomrule
\end{tabular}
\end{table}

The maximum absolute weight deviation across all cells and components is
0.032 (SSM weight, LOO b1c5 fold), and the mean absolute deviation is
0.005.
The component rank order (WD $>$ RCVAE $>$ Defl $>$ HT2 $>$ SSM by weight)
is preserved in 6 of 7 LOO folds; the single exception is the b1c5 fold,
where SSM and RCVAE swap ranks 2 and 5 while the top-two (WD, RCVAE/SSM)
and top-1 (WD) positions are stable.
These results indicate that the RWMHI weight allocation is stable under
single-cell removal and that the reported alarm cycles are unlikely to change
materially under LOO weight perturbation.

\subsection*{Ablation Study}
\label{app:ablation}

Table~\ref{tab:ablation} presents a formal ablation comparing individual
detectors, bundle medians, and the RWMHI cross-bundle fusion at $f_c = 0.10$.
Detection rate is the fraction of the seven cells in which a CUSUM alarm
was raised; lead time is the cell's end-of-life cycle minus the first-alarm
cycle, averaged over detected cells.

\begin{table}[h]
\centering
\caption{Ablation at $f_c = 0.10$ ($n = 7$ cells).
  \emph{Det\,\%}: fraction of cells with a raised alarm.
  \emph{Median}: median first-alarm cycle across detected cells.
  \emph{IQR}: interquartile range.
  \emph{Mean lead}: mean lead time (cycles before 80\,\% capacity threshold).
  Isolation Forest (29\,\% detection at $f_c = 0.10$) and
  Reliability-Weighted MHI are shown separately; all others are grouped
  by signal class.
  Bundle medians use the class-level median (all detections pooled).
  Three cells (NASA B0005, B0006, MATR b1c20) have $N_{\mathrm{comm}} <
  60$ cycles and are therefore floor-constrained (Section~\ref{sec:cusum});
  the RWMHI IQR lower bound of 62 cycles and the pooled mean lead times
  are partly influenced by this floor.}
\label{tab:ablation}
\small
\begin{tabular}{lrrrr}
\toprule
Method & Det\,\% & Median & IQR & Mean lead \\
\midrule
\multicolumn{5}{l}{\emph{Complexity-entropy bundle}} \\
\quad SVD Compression Entropy       & 86\% &  84 & [66--107]  & 249 \\
\quad Information Evolution         &100\% &  81 & [67--167]  & 213 \\
\quad Mutual Information            & 71\% &  78 & [77--108]  & 223 \\
\quad \emph{Bundle median}          &100\% &  80 & [66--110]  & 250 \\
\addlinespace
\multicolumn{5}{l}{\emph{Magnitude-shift bundle}} \\
\quad Mahalanobis Deflation         &100\% & 110 & [86--145]  & 217 \\
\quad Hotelling's $T^2$             &100\% & 110 & [66--144]  & 232 \\
\quad Window Distance               &100\% &  95 & [61--101]  & 257 \\
\quad Wasserstein Distance          &100\% & 114 & [83--128]  & 233 \\
\quad Sliced Wasserstein Distance   &100\% & 122 & [83--154]  & 219 \\
\quad \emph{Bundle median}          &100\% & 105 & [64--136]  & 257 \\
\addlinespace
\multicolumn{5}{l}{\emph{Predictive-residual bundle}} \\
\quad VAE Reconstruction            & 71\% & 189 & [99--231]  & 175 \\
\quad RC-VAE Reconstruction         &100\% & 104 & [84--124]  & 235 \\
\quad SSM Prediction Error          &100\% & 125 & [66--205]  & 198 \\
\quad GRU-CausalSSM Residuals       &100\% & 101 & [90--119]  & 240 \\
\quad \emph{Bundle median}          &100\% & 106 & [84--151]  & 252 \\
\addlinespace
\multicolumn{5}{l}{\emph{Excluded (low reliability)}} \\
\quad Isolation Forest              & 29\% & --- & ---        & --- \\
\addlinespace
\multicolumn{5}{l}{\emph{Cross-bundle fusion}} \\
\quad \textbf{RWMHI (5 components)} &\textbf{100\%} & \textbf{95} & \textbf{[62--102]} & \textbf{257} \\
\bottomrule
\end{tabular}
\end{table}

The ablation shows that all five individual RWMHI components achieve
100\,\% detection, justifying their selection.
VAE (71\,\%) and Isolation Forest (29\,\% at $f_c = 0.10$) are correctly excluded.
The RWMHI median first-alarm (95 cycles) is earlier than both the
Magnitude bundle median (105 cycles) and the Predictive bundle median
(106 cycles), confirming a modest but consistent fusion gain.
The Complexity bundle achieves the earliest pool-level median (80 cycles)
but at the cost of 29\,\% missed detections for two of its three components;
cross-cell reliability is a stronger operational requirement than early firing.

\end{document}